\documentclass[aps,prd,reprint, superscriptaddress, nofootinbib, floatfix]{revtex4-2}
\usepackage{graphicx, subfigure}        % Including figure files
\usepackage{aas_macros} 
\usepackage{amssymb,amsmath,placeins,epsfig,array,bm, mathtools}
\usepackage[usenames,dvipsnames]{color}
\usepackage[normalem]{ulem}
\usepackage[breaklinks,colorlinks,urlcolor=blue,citecolor=blue,linkcolor=blue]{hyperref}
\usepackage{float}
\usepackage[T1]{fontenc}
\usepackage{times}
\usepackage{soul}
\usepackage{color, colortbl}

\newcommand{\dif}{\mathop{}\!{d}}

\newcommand{\bol}[1]{\boldsymbol{#1}}

\DeclareMathOperator{\Loss}{\mathcal{L}}

\DeclareMathOperator{\x}{\mathbf{x}}

\DeclareMathOperator{\p}{\mathrm{p}}

\DeclarePairedDelimiterX{\infdivx}[2]{(}{)}{#1\;\delimsize\|\;#2}

\begin{document}

%Title of paper
\title{Deep learning insights into cosmological structure formation}

\author{Luisa Lucie-Smith}
\email[]{luisals@mpa-garching.mpg.de}
\affiliation{Max-Planck-Institut für Astrophysik, Karl-Schwarzschild-Str. 1, 85748 Garching, Germany}
\affiliation{Department of Physics \& Astronomy, University College London, Gower Street, London WC1E 6BT, UK}

\author{Hiranya V. Peiris}
\affiliation{Department of Physics \& Astronomy, University College London, Gower Street, London WC1E 6BT, UK}
\affiliation{The Oskar Klein Centre for Cosmoparticle Physics, Department of Physics, Stockholm University, AlbaNova, Stockholm, SE-106 91, Sweden}

\author{Andrew Pontzen}
\affiliation{Department of Physics \& Astronomy, University College London, Gower Street, London WC1E 6BT, UK}

\author{Brian Nord}
\affiliation{Fermi National Accelerator Laboratory, P.O. Box 500, Batavia, IL 60510, USA}
\affiliation{Department of Astronomy and Astrophysics, University of Chicago, Chicago, IL 60637, USA}
\affiliation{Kavli Institute for Cosmological Physics, University of Chicago, Chicago, IL 60637, USA}

\author{Jeyan Thiyagalingam}
\affiliation{Scientific Computing Department, Rutherford Appleton Laboratory, Science and Technology Facilities Council, Harwell Campus, Didcot, OX11 0QX}

\date{\today}

% Keywords are not mandatory, but authors are strongly encouraged to provide them. If provided, please include two to five keywords, separated by the pipe symbol, e.g:
%\keywords{Keyword 1 $|$ Keyword 2 $|$ Keyword 3 $|$ ...} 

\begin{abstract}
The evolution of linear initial conditions present in the early universe into extended halos of dark matter at late times can be computed using cosmological simulations. However, a theoretical understanding of this complex process remains elusive; in particular, the role of anisotropic information in the initial conditions in establishing the final mass of dark matter halos remains a long-standing puzzle. Here, we build a deep learning framework to investigate this question. We train a three-dimensional convolutional neural network (CNN) to predict the mass of dark matter halos from the initial conditions, and quantify in full generality the amounts of information in the isotropic and anisotropic aspects of the initial density field about final halo masses. We find that anisotropies add a small, albeit statistically significant amount of information over that contained within spherical averages of the density field about final halo mass. However, the overall scatter in the final mass predictions does not change qualitatively with this additional information, only decreasing from 0.9 dex to 0.7 dex. Given such a small improvement, our results demonstrate that isotropic aspects of the initial density field essentially saturate the relevant information about final halo mass. Therefore, instead of searching for information directly encoded in initial conditions anisotropies, a more promising route to accurate, fast halo mass predictions is to add approximate dynamical information based e.g. on perturbation theory. More broadly, our results indicate that deep learning frameworks can provide a powerful tool for extracting physical insight into cosmological structure formation.
\end{abstract}

% insert suggested keywords - APS authors don't need to do this
%\keywords{}

%\maketitle must follow title, authors, abstract, and keywords
\maketitle

\section{Introduction}
The formation of cosmic structures in the Universe is driven by the gravitational collapse of initially small perturbations in the density of matter, which grow over time into extended halos of dark matter. Computer simulations are the most accurate method available to compute the non-linear evolution of dark matter over cosmic time \cite{Efstathiou1985, Jenkins2001, Navarro1997, gadget}.  Given the initial conditions and a cosmological model, $N$-body simulations follow the evolution of particles governed by the laws of gravity. Despite being able to compute the evolution of matter in the Universe, simulations alone do not provide a straightforward answer to how dark matter halos acquire their characteristic properties -- such as mass, shape, inner profile and spin -- from the initial density perturbations. 

On the other hand, analytic theories of structure formation can provide a qualitative understanding of the connection between the early- and late-time Universe. By construction, all analytic frameworks present a far more simplified view of gravitational evolution relative to solving N-body dynamics, and therefore sacrifice some predictive accuracy but allow for a much clearer interpretation. Spherical collapse models provided us with the widely-accepted idea that spherical overdensities encode the primary information about halo collapse \cite{PressSchechter1974, Bond1991}. Ellipsoidal collapse models yield a significant improvement over spherical collapse ones in predicting statistical quantities of the large-scale structure of the Universe, such as the halo mass function \cite{Doroshkevich1970, Bond&Myers1996, ShethTormen1999, ShethMoTormen2001, ShethTormen2002}. Ellipsoidal collapse models do not directly use anisotropic features of the field in reaching their conclusion; instead, the models introduce free parameters within the spherical collapse framework, motivated by arguments about tidal shear effects. Those parameters are then fitted to numerical simulations. Therefore, whether or not anisotropic features of the initial density field have a role in establishing final halo masses remains a long-standing question in cosmological structure formation.

In previous work \cite{LucieSmith2018, LucieSmith2019}, we proposed a novel approach based on machine learning to gain new insights into physical aspects of the early Universe responsible for halo collapse. The approach consists of training a machine learning algorithm to learn the relationship between the early universe and late-time halo masses directly from numerical simulations. The learning of the algorithm is based on a set of inputs, known as \textit{features}, describing pre-selected physical aspects about the linear density field in the initial conditions. We trained the algorithm on spherical overdensities (motivated by spherical collapse models) and tidal shear information (motivated by ellipsoidal collapse models) in the local environment surrounding each dark matter particle in the initial conditions. Contrary to existing interpretations of the Sheth-Tormen ellipsoidal collapse model \cite{ShethMoTormen2001, ShethTormen2002}, we found that the addition of tidal shear information does not yield an improved model of halo collapse compared to a model based on density information alone \cite{LucieSmith2018, LucieSmith2019}. This approach is limited by the need to explicitly construct a set of informative features, which relies on simplified analytic approximations of halo collapse. Due to this limitation, our previous work tackled a limited science question: the role of one specific anisotropic feature of the initial conditions (the tidal shear) in predicting final halo masses.

In this work, we extend our approach to a deep learning framework based on convolutional neural networks (CNNs) \cite{LeCun2015, Bengio2009deep}. 
Unlike standard machine learning algorithms or analytic descriptions, CNNs do not require specification of pre-selected features from the data; instead, they are trained to extract information directly from raw data. This framework allows us to address a major long-standing issue in cosmology: the role of all anisotropic aspects of the initial density field in establishing final halo masses.

The structure of a CNN is closely similar to that of existing analytic halo collapse descriptions: the information in the initial conditions is compressed into a set of features, which are then combined in a non-linear way to provide a halo mass prediction. Our CNN approach therefore yields a simplified description of halo collapse, and should not be expected to provide perfect predictive accuracy relative to detailed N-body simulations. However, since a CNN allows for the extraction of arbitrary features, it yields a model of halo collapse that far transcends the capabilities of current analytic approaches while following the same basic setup. Our approach can therefore be seen as a generalization of existing analytic approaches, which are limited to the extraction of spherical features from the initial conditions.

In Sec.~\ref{sec:overview}, we present an overview of our deep learning framework, followed by a description of the simulated data used to train the machine learning model and details on the CNN's architecture. We present the halo mass predictions returned by the model in Sec.~\ref{sec:predictions}, comparing to expectations from previous work. We then move on to interpreting the learnt mapping between initial conditions and halo mass in Sec.~\ref{sec:averaged}, by testing the impact on the model's performance as we remove from the inputs certain physical aspects of the initial conditions. Finally, we test the robustness of our model in Sec.~\ref{sec:z0}, and conclude with a discussion on the implications of our work in Sec.~\ref{sec:conclusions}.

\section{The deep learning framework}
\label{sec:overview}

\begin{figure*}[t]
\centering
	\includegraphics[width=0.9\textwidth]{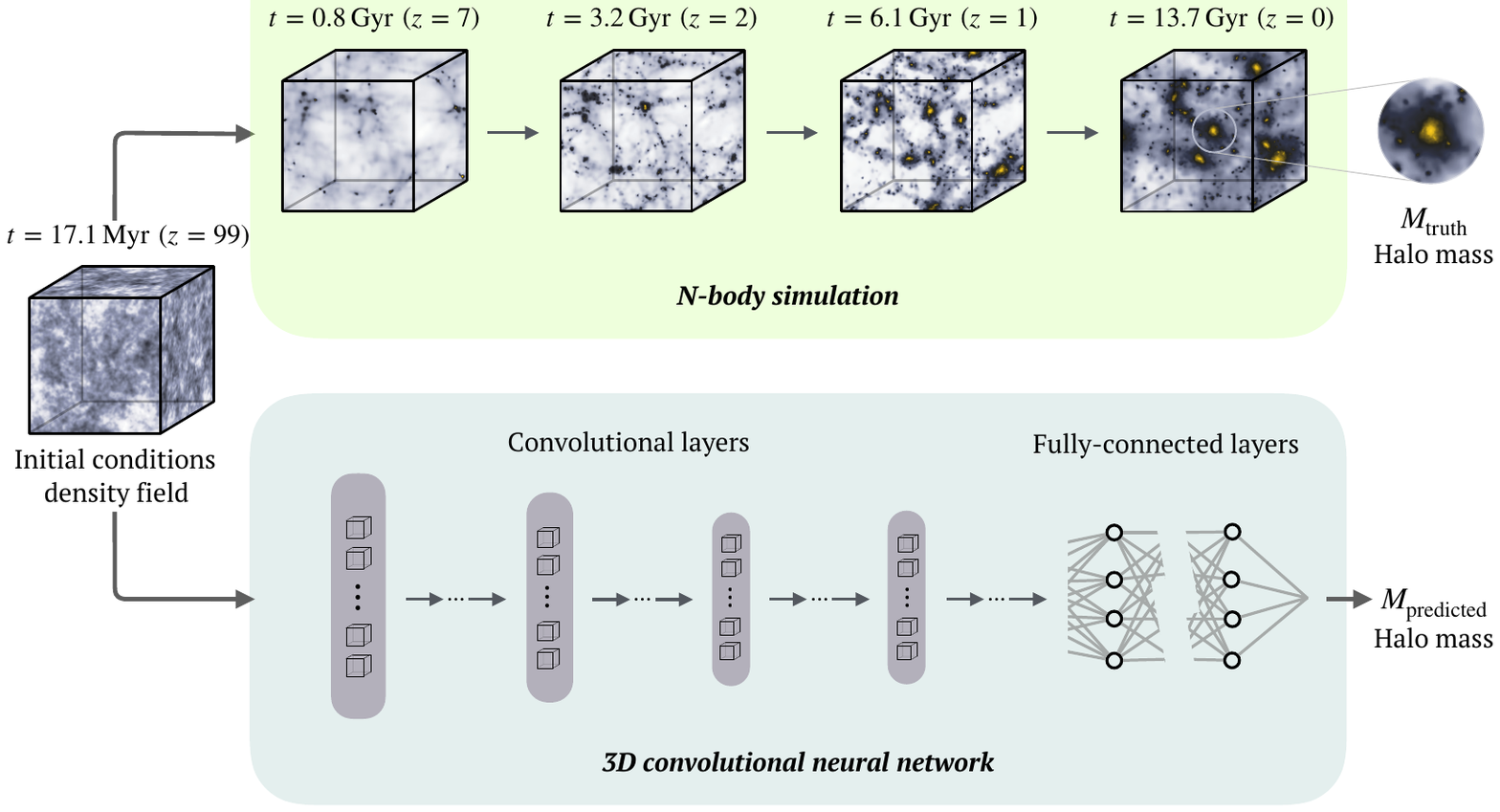}
    \caption{$N$-body simulations of cosmological structure formation can accurately compute the gravitational evolution of dark matter over cosmic time, but do not provide a physical understanding of how cosmic structures arise from the initial conditions. We train a CNN model to learn the relationship between the initial density field and the final dark matter halos, given examples from $N$-body simulations. The inputs to the CNN are given by the initial density field surrounding each dark matter particle and the outputs are the mass of the dark matter halos to which each particle belongs at $z=0$. The aim is to interpret the mapping learnt by the CNN in order to gain physical insights into dark matter halo formation.}
    \label{fig:dlhalos_diagram}
\end{figure*}

$N$-body simulations are the most general and accurate available approach to compute the clustering evolution of matter at all scales (Fig.~\ref{fig:dlhalos_diagram}). Simulations start from early times, when the Universe was filled with small matter density perturbations that are described by a Gaussian random field. The simulations then follow the evolution of these fluctuations as they enter the non-linear regime, through the emergence of self-gravitating dark matter halos wherein galaxies form. The final product of the simulation resembles the Universe we observe today, characterized by dark matter halos embedded in a web of filamentary large-scale structure.

Our aim is to develop a deep learning framework that can be used to learn about the physical connection between the early Universe and the mass of the final dark matter halos (Fig.~\ref{fig:dlhalos_diagram}). We focus on the mass of dark matter halos as it is the halos' primary characteristic, but our framework can similarly be applied to other halo properties. The CNN is trained to predict the final mass of the halo to which any given dark matter particle in the simulation belongs at the present time. The input to the CNN for a given particle is given by the initial density field in a cubic sub-region of the initial conditions of the simulation, centered on the particle's initial position. The CNN is therefore trained to learn a particle-by-particle mapping from the initial conditions to the final mass of the halo to which each particle belongs.
Our results are insensitive to the simulation box size and the choice of whether to use overdensity or gravitational potential as input, as discussed in Appendix \ref{sec:potential}.

In cosmology, deep learning has become a popular method to learn mappings that require computationally expensive $N$-body simulations. Examples of CNN applications to simulations include estimating cosmological parameters from the dark matter or galaxy distribution \cite{Ravanbakhsh2017, Mathuriya2018, Pan2020, Villaescusa-Navarro2020, Ntampaka2020, Moster2020}, generating higher-resolution versions of low-resolution $N$-body simulations \cite{KodiRamanah2020, Li2021}, as well as emulating the mapping between Zel'dovich-displaced and non-linear density fields \cite{He2019}, or that between the dark matter and galaxy distributions \cite{Zhang2019, Moster2020}. The models are evaluated on global summary statistics such as two-point or three-point correlation functions \cite{Ramanah2019, Mathuriya2018, Ntampaka2020, Ravanbakhsh2017, Zhang2019}. Our work differs from such applications primarily in the aim: it is not to develop fast $N$-body surrogates, but rather to make use of deep learning to gain physical insight into the formation of cosmic structures within the simulations. Our CNN model returns particle-specific predictions, yielding a halo collapse model that can describe the non-linear evolution of the density field from any initial location in the simulation.

\subsection{Simulations}
We generated the training data from 20 dark matter-only $N$-body simulations produced with \texttt{P-GADGET-3} \cite{gadget2, gadget}, each consisting of a box of size $L= 50 \, \mathrm{Mpc} \, h^{-1}$ (comoving) and $N = 256^3$ simulation particles evolving from $z=99$ to $z=0$. We tested the impact of the simulation box size by repeating the analysis using three simulations with box size $L= 200 \, \mathrm{Mpc} \, h^{-1}$ and the same mass resolution ($N = 1024^3$). We found no significant change in the final predictions of the model, demonstrating that a box of size $L= 50 \, \mathrm{Mpc} \, h^{-1}$ is sufficiently large to capture the relevant environmental effects for the halo population considered in this analysis. The results shown in the paper are for models trained on the $L= 50 \, \mathrm{Mpc} \, h^{-1}$ cosmological boxes. We made use of \texttt{pynbody} \cite{pynbody} to analyse the information contained in the simulation snapshots. The simulations adopt a WMAP$5$ $\Lambda$CDM cosmological model; the cosmological parameters are given by $\Omega_{\Lambda} = 0.721$, $\Omega_{\mathrm{m}} = 0.279$, $\Omega_{\mathrm{b}} = 0.045$, $\sigma_{8} = 0.817$, $h = 0.701$ and $n_s = 0.96$ \cite{WMAP}. Some of the simulations are part of a suite of existing simulations, which were performed at times where these cosmological parameters were up-to-date; the newer simulations were then run with the same set of old parameters for consistency. However, we do not expect our conclusions to change when updating the cosmological parameters to more recent constraints from observations \cite{Aghanim2020}, as demonstrated in previous work \cite{LucieSmith2018}. Each simulation is based on a different realization of a Gaussian random field drawn from the initial power spectrum of density fluctuations, generated using \texttt{genetIC} \cite{Stopyra2020}. The simulation particles of the validation and testing data were randomly drawn from 4 additional, independent simulations to those used for training and their inputs/outputs were generated in the same way as for the training data.

Dark matter halos were identified at $z=0$ using the \texttt{SUBFIND} halo finder \cite{gadget}, a friends-of-friends method with a linking length of $0.2$, with the additional requirement that particles in a halo be gravitationally bound. We consider the entire set of bound particles that make up a halo and do not account for substructure within halos. The resolution and volume of the simulation limit the resulting range of halo masses: the lowest-mass halo has $M=2.6\times10^{10}~\mathrm{M}_{\odot}$ and the highest $M=4.1\times10^{14}~\mathrm{M}_{\odot}$. We restrict our analysis even further to the mass range $\log \left( M/\mathrm{M}_{\odot} \right) \in \left[ 11, 13.4 \right]$. This is because halos with mass $M \lesssim 10^{11} \mathrm{M_\odot}$ contain less than $\sim 100$ particles and are therefore not well resolved in the simulation, whereas halos with mass $M \gtrsim 3 \times 10^{13} \, \mathrm{M_\odot}$ are under-represented as a result of the small volume of our simulations.

\subsection{Inputs and outputs of the deep learning models}
The training data consist of dark matter particles randomly drawn from the ensemble of particles in the simulations which belong to dark matter halos at $z=0$; we consider all particles within halos, and not just those at the halo centres.

The final snapshots of the simulations ($z = 0$) were used to label each dark matter particle with its ground truth variable, given by the logarithmic mass of the dark matter halo to which each dark matter particle belongs. We only consider particles that make up dark matter halos at $z=0$ in this analysis. The ground truths were rescaled to the range $\left[-1, 1 \right]$ before training; this step sets a similar scale and dynamic range for the inputs and outputs of the model, which facilitates the model's training.

The density field in the initial conditions of the simulations ($z = 99$) was used to generate the deep learning inputs associated with each particle. In the initial conditions, the density field is given by a random realization $\delta \left( \mathbf{x}, t_\mathrm{initial} \right)$ on a uniform $256^3$ grid in the $(50 \, \mathrm{Mpc} \, h^{-1})^3$ simulation volume. The input associated with any given particle is given by $\delta \left( \mathbf{x}, t_\mathrm{initial} \right)$ in a cubic sub-region of the full simulation centered on the particle's initial position. The density at every voxel of the cubic box is estimated from the positions of the particles in the initial conditions; specifically, we estimate the density at the location of each particle following an SPH procedure where the SPH kernel smoothing length depends on each particle's 32 nearest neighbours. Our results are insensitive to the exact number of nearest neighbours. This sub-volume has size $L= 15 \, \mathrm{Mpc}\, h^{-1}$ (comoving) and resolution $N=75^3$. 

The size of the sub-box was chosen to be large enough to capture large-scale information that is relevant to the algorithm to learn the initial conditions-to-halo mass mapping. In previous work, we trained a different machine learning algorithm to infer final halo masses in the same mass range based on pre-computed features of the initial conditions density field \cite{LucieSmith2019}. We found that the machine learning model was able to learn relevant information from the smoothed density field up to a scale of $M_\mathrm{smoothing} \sim 10^{14} \, \mathrm{M}_\odot$. Therefore, we chose a sub-box length $L_\mathrm{box} = 15\, \mathrm{Mpc} \, h^{-1}$, which encloses a total mass of $M \sim 4 \times 10^{14} \,\mathrm{M}_\odot$, which is more than the largest relevant mass scale adopted in our previous work. We found that increasing the volume did not change the performance of the network. On the other hand, a smaller volume would lead to degradation in the predictions of particles in high-mass halos. The resolution of the sub-box was chosen such that the length of each voxel, $l_\mathrm{voxel}$, is the same as the initial grid spacing in the simulation i.e., $l_\mathrm{voxel} = 0.2 \, \mathrm{Mpc} \, h^{-1}$ (comoving). This is the highest possible choice of resolution. The training set inputs were rescaled to have 0 mean and standard deviation 1; the same rescaling was then applied to the validation and test sets. 

\subsection{The deep learning model}
The deep learning model consists of a 3D CNN, made of six convolutional layers and three fully-connected layers. Although CNNs are generally applied to two-dimensional images, we used three-dimensional kernels in the convolutional layers that can be applied to the 3D initial density field of the $N$-body simulation. The convolutions were performed with 32, 32, 64, 128, 128, 128 kernels for the six convolutional layers, respectively. The kernels have size $3\times3\times3$. All convolutional layers (but the first one) are followed by max-pooling layers; their output is then used as input to the non-linear leaky rectified linear unit (LeakyReLU) \cite{Nair2010} activation function. We refer the reader to Appendix \ref{sec:architecture} for more details on the CNN architecture. By training the network across many examples of particles across many simulations, the model learns to identify the aspects of the initial density field which impact the final mass of the resulting halos.

Training the deep learning model requires solving an optimization problem. The parameters of the model, $ \bol{w}$, are optimized to minimize the loss function, $\Loss_w (\bol{M}_\mathrm{true}, \bol{M}_\mathrm{pred})$, which measures how closely the predictions, $\bol{M}_\mathrm{pred}$, are to their respective ground truths, $\bol{M}_\mathrm{true}$, for the training data. The model consists of a large number of parameters, thus making it highly flexible. As a result, CNNs are often prone to overfitting the training data, without generalizing well to unobserved test data. To overcome this, regularization techniques are employed by incorporating an additional penalty term into the loss function. We designed a custom loss function given by
\begin{gather}
\Loss_w (\bol{M}_\mathrm{true}, \bol{M}_\mathrm{pred}) = \Loss_\mathrm{pred} (\bol{M}_\mathrm{true}, \bol{M}_\mathrm{pred}) + \Loss_\mathrm{reg}(\bol{w}),
\label{eq:loss_overview}
\end{gather}
where $ \Loss_\mathrm{pred} (\bol{M}_\mathrm{true}, \bol{M}_\mathrm{pred})$ is the predictive term, measuring how well the predicted values match the true target values, and $\Loss_\mathrm{reg}(\bol{w})$ is the regularization term. The predictive term can be re-expressed as $ \Loss_\mathrm{pred} = - \ln \left[ p \left( \bol{M}_\mathrm{true}, \mid  \bol{w}, \mathcal{M} \right) \right]$, where $p \left(\bol{M}_\mathrm{true} \mid  \bol{w}, \mathcal{M} \right)$ describes the the probability distribution of ground truth values $\bol{M}_\mathrm{true}$, given the predicted values $\bol{M}_\mathrm{pred}$ of the training data returned by the 3D CNN model $\mathcal{M}$ with parameters $\bol{w}$. A common choice in the community is that of a Gaussian or Laplacian distribution, yielding the popular mean-squared-error or mean-absolute-error losses for regression. We found that a Cauchy distribution provides a better description of the data, as it contains broader tails than those of a Gaussian distribution. The scale parameter $\gamma$ of the Cauchy distribution, which specifies the half-width at half-maximum, was optimized via back-propagation \cite{Rumelhart1986} during the training procedure of the CNN, similar to the way the model parameters $\bol{w}$ are optimized. The regularization term in \eqref{eq:loss_overview}, $\Loss_\mathrm{reg}(\bol{w})$, was designed to simultaneously (i) improve the optimization during training by preventing the algorithm from overfitting the training data and (ii) compress the neural network model into the smallest number of parameters without loss in performance. We refer the reader to Appendix \ref{sec:loss} for more details on our custom loss function.

The parameters $\bol{w}$ were optimized during training via back-propagation, which consists of the chain rule for partial differentiation applied to the gradient of the loss with respect to the parameters. The training proceeds for thousands of iterations, each consisting of a forward and a backward pass. In the forward pass, the input runs through the network and reaches the output layer, and in the backward pass, the parameters of the network are updated to minimize the loss function evaluated for the training data. The algorithm was trained on $200$,$000$ particles, randomly drawn from the ensemble of particles of $20$ simulations based on different realizations of the initial conditions; it was then validated using 10,000 particles randomly drawn from a single simulation, and tested on 99,950 particles drawn from four additional independent simulations. The training set was sub-divided into batches, each made of $64$ particles. Batches were fed to the network one at a time, and each time the CNN updates its parameters according to the samples in that batch. Training was done using the \texttt{AMSGrad} optimizer \cite{Reddi2019}, a variant of the widely-used \texttt{Adam} optimizer \cite{Kingma2014}, with a learning rate of 0.00005. The learning rate was optimized via cross-validation, together with $\alpha$, the parameter weighting the regularization term in the loss function. Early stopping was employed to interrupt the training at the epoch where the validation loss reaches its minimum value.

\section{Predicting the mass of dark matter halos from the initial conditions}
\label{sec:predictions}

\begin{figure*}
\centering
	\includegraphics[width=\textwidth]{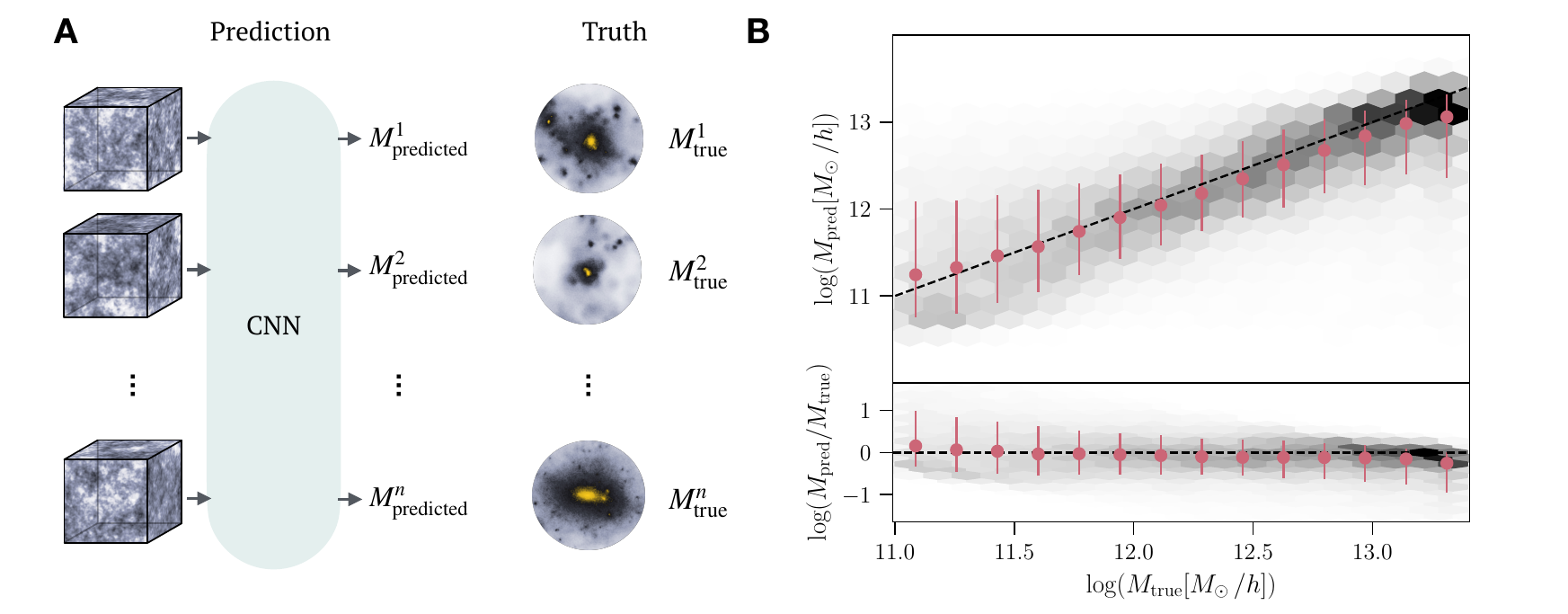}
    \caption{(\textbf{A}) The CNN makes predictions for simulation particles that occupy different regions of the initial conditions of the simulation. These particles end up in halos which differ not only in their mass, but also in their formation history, large-scale environment, and amount of sub-structure within the halos. The CNN must identify from the initial density field the features that impact the final mass of the resulting halos. (\textbf{B}) Halo mass predictions returned by a CNN trained on the initial density field surrounding each dark matter particle's initial position. The predictions are shown against the ground truth halo mass values as a two-dimensional histogram in the top panel, while the bottom panel shows the residuals $\log(M_{\rm pred}/M_{\rm true})$. The errorbars in the top (bottom) panel show the median and 68\% confidence interval of the predictions (residuals) in bins of ground-truth mass values.}
    \label{fig:CNN_z99}
\end{figure*}

We applied our trained CNN model to particles from independent simulations not used for training. The CNN predicts the mass of the halo to which the particles will belong at $z=0$. The test set contains particles belonging to randomly-selected dark matter halos with mass $\log \left( M/\mathrm{M_\odot}  \right) \in [11, 13.4]$ (Fig.~\ref{fig:CNN_z99}A). This mass range is set by the resolution and volume of our simulations; halos of mass $\log \left( M/\mathrm{M_\odot}  \right) \lesssim 11$ are not well resolved and those with mass $\log \left( M/\mathrm{M_\odot}  \right) \gtrsim 13.4$ are rare (and therefore under-represented) in the small volume of our simulations. The halos in the test set do not only differ in mass; they also differ by factors such as their formation history and their large-scale environment, which contribute significantly to making the mapping between initial conditions and halo masses challenging. For example, halos of the same mass may have assembled smoothly through small accretion events or violently through mergers with other massive structures; they may have formed in isolated regions of the Universe or close to filaments and other massive objects. Particles that belong to the same halo may even have experienced significantly different dynamical histories: those in the inner region of halos are more likely to have been bound to the proto-halo patch from very early times, whereas those in the outskirts may have been more recently accreted onto the halo through late-time halo mergers, tidal stripping or accretion events. All this variability in the formation process of dark matter halos is not explicitly presented to the deep learning model; the CNN is faced with the task of finding features in the initial conditions which contain information about the complex, non-linear evolution of halos. 

Our problem setup is closely related to that of analytic models: features are extracted via convolutions from the initial conditions and combined non-linearly to yield a halo mass prediction. Our CNN approach provides a major generalization over analytic models since the former is capable of extracting any arbitrary form of features from the inputs. Thus, we expect the CNN model to return halo mass predictions that are at least as accurate as those of state-of-the-art analytic approximations, or more accurate if there exists additional features of the initial conditions beyond those captured by analytic models that yield an improved description of halo collapse.

We compare the predictions made by the CNN to the true halo masses of the test set particles (Fig.~\ref{fig:CNN_z99}B). The top panel shows the predicted against the true halo mass values as a two-dimensional histogram, while the bottom panel shows the residuals $\log(M_{\rm pred}/M_{\rm true})$. The errorbars in the top (bottom) panel show the median and 68\% confidence interval of the predictions (residuals) in bins of ground-truth values.
The black dashed line shows $y=x$ and represents the idealized case of 100\% accuracy. The predictions' distributions are characterized by large variances and skewness throughout the whole mass range of halos, although the maxima of the posterior distributions are in the correct location. The variance in the distributions is larger for low-mass halos compared to high-mass halos. The extent of the variance in the predictions of the CNN is consistent with expectations from analytic models, such as the Sheth-Tormen ellipsoidal collapse model \cite{ShethMoTormen2001}, which model a similar mapping between pre-selected features of the initial conditions and halo mass. We show and quantify the comparison between the CNN predictions and those of analytic models in Appendix \ref{sec:analytic}.

\begin{figure*}
\centering
\includegraphics[width=\textwidth]{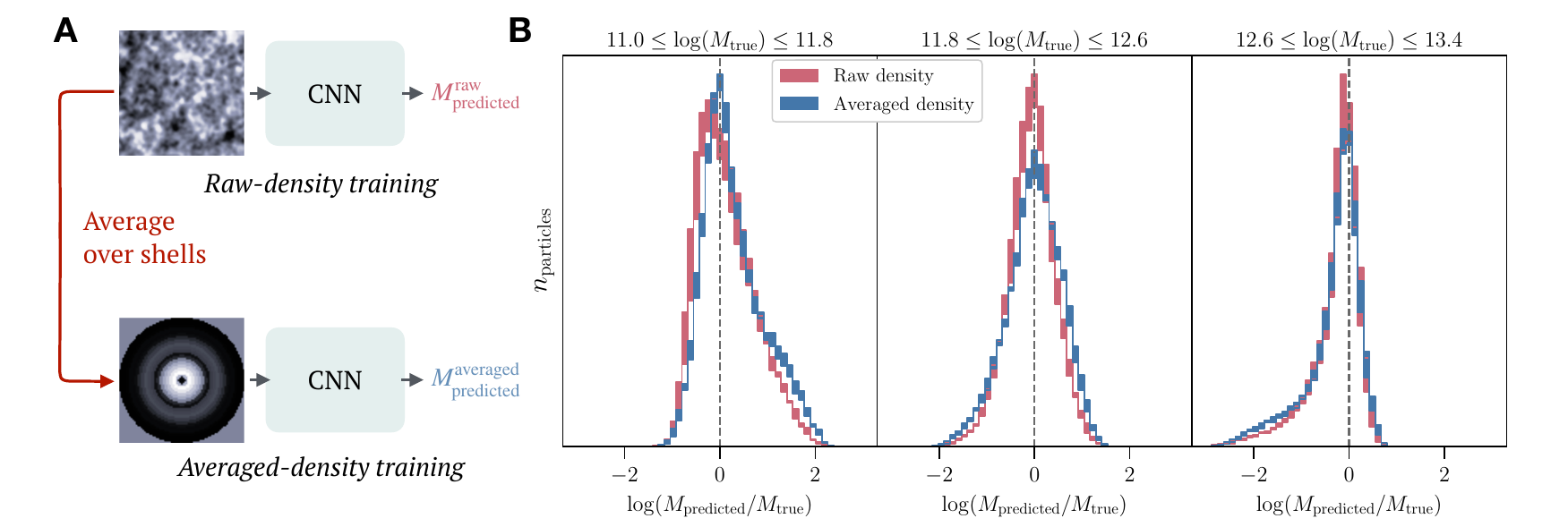}
\caption{(\textbf{A}) We re-train the model on inputs where the density in the initial conditions is averaged over shells so that any anisotropic information is removed. The two models each return a set of predictions for the test set particles. (\textbf{B}) We compare the predictions returned by the \textit{raw-density} training set model and the \textit{averaged-density} training set model. The histograms show the difference between the predicted and the true log halo mass for particles split into three mass bins of halos. The bands of the histograms capture the scatter in the predictions of each model trained with four different random seeds. The two models show similar residual distributions, except for a slightly smaller variance in the residual distribution of particles in the mid-mass range of halos.}
\label{fig:raw_vs_averaged}
\end{figure*}

\section{Interpreting the information learnt by the CNN}
\label{sec:averaged}
Our goal it to interpret the mapping learnt by the deep learning model to understand which aspects of the initial conditions the CNN extracts to make its predictions. In particular, we wish to test whether anisotropic aspects of the density field play a major role in establishing final halo mass.

To do this, we modified the inputs to the CNN to remove any anistropic information about the 3D density field and re-trained the CNN. Given 20 concentric shells around the center of the box, the inputs were constructed by assigning to each voxel within a given shell the average density within that shell (Fig.~\ref{fig:raw_vs_averaged}A). The concentric shells were evenly spaced in radius $r$ within the range $r \in \left[ 2, \, 75 \right]$ voxels. Outside the largest shell that fits entirely within the box, voxels were assigned the average density within those parts of the concentric shell that intersect with the box. This ensured that no additional information from outside the input sub-box region was used to construct the inputs. This procedure implies that each voxel of the 3D input sub-box only carries information about the spherically-averaged density. We call this model the \textit{averaged-density} training set model, whereas the original one which uses the full initial density field as input is denoted as the \textit{raw-density} training set model. The two separately-trained models were used to return their own halo mass predictions for the same set of test particles. 

Figure \ref{fig:raw_vs_averaged}B shows the comparison between the predictions of the two models, one trained on the raw initial density field and one trained only on spherically-averaged information. The bands of the histograms capture the scatter in the predictions of each model when trained using five different random seeds. We find that the two models return qualitatively similar predictions in the halo mass range $11 \leq \log(M/\mathrm{M_\odot}) \leq 13.4$, for the same set of test particles. 

\begin{figure}
\centering
\includegraphics[width=\columnwidth]{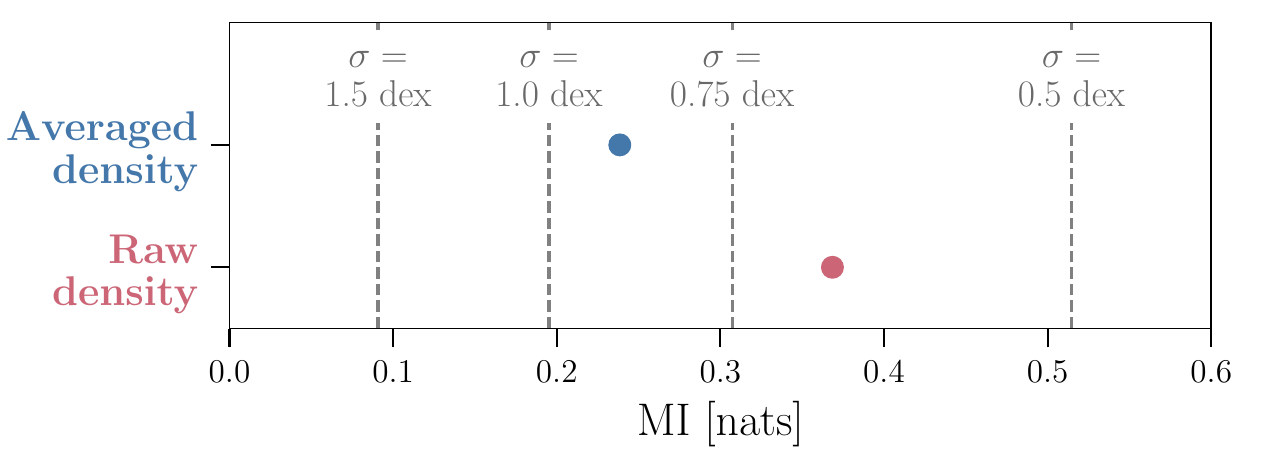}
\caption{Mutual information between predicted and ground truth halo mass values for the raw-density and averaged-density models. The horizontal grey lines show the value of the MI for mock predicted halo mass values constructed by adding Gaussian noise to the ground truth values with standard deviation of 1.5, 1., 0.75, 0.5 dex, respectively.}
\label{fig:raw_vs_averaged_MI}
\end{figure}

To quantify the similarity between the predictions of the two models, we used the information-theoretic metric of mutual information (MI) between the predicted and ground truth halo mass values of the test set particles. In contrast to linear correlation measures such as the $r$-correlation, MI measures the full (linear and non-linear) dependence between two variables. This allowed us to quantify and compare the amount of information captured by each model about the ground truth halo masses.

Mathematically, the MI between two continuous variables $X$ and $Y$ with values over $\mathcal{X} \times \mathcal{Y}$, $I (X, Y)$ is defined as:
\begin{align}
    I (X, Y) \equiv \int_{\mathcal{X} \times \mathcal{Y}}  p_{(X, Y)}(x, y) \ln{\frac{p_{(X, Y)}(x, y)}{ p_X(x) p_Y(y)}} \dif x \dif y \ ,
    \label{eq:MI}
\end{align}
where $p_{(X, Y)}$ is the joint probability density distribution of $X$ and $Y$, and $p_X$ and $p_Y$ are their marginal distributions, respectively. MI as defined by Eq.~\eqref{eq:MI} is measured in natural units of information (nats). The MI was estimated using the publicly-available software GMM-MI \cite{Piras2022}, which performs density estimation using Gaussian mixtures to estimate $p_{(X, Y)}$ in Eq.~\eqref{eq:MI} and provides MI uncertainties through bootstrap. \footnote{The distribution of ground truth values contains two hard boundaries at $\log (M_\mathrm{min}/[\mathrm{M}_\mathrm{sol}\, h^{-1}])=11$ and $\log (M_\mathrm{max}/[\mathrm{M}_\mathrm{sol}\, h^{-1}])=13.4$, which can be problematic for the Gaussian mixture density estimation of $p_{(X, Y)}$. Moreover, the fact that there exists fewer halos at the high mass end also introduces some discretization to the distribution of ground truth values at high mass. To correct for these effects, we add a small random noise and then apply an arctanh transformation to the ground truth values. Since mutual information is invariant under invertible, non-linear transformations and the added noise is smaller than the discretization, this does not affect the value of the MI, but makes the density estimation procedure more robust.}

We find a value of $I(M_\mathrm{pred},M_\mathrm{truth}) =0.370 \pm 0.004$ for the raw density model and $I(M_\mathrm{pred},M_\mathrm{truth})=0.240 \pm 0.002$ for the averaged density one, as shown in Fig.~\ref{fig:raw_vs_averaged_MI}. The scatter in the MI between predicted and true halo mass values of the same model initiated with different random seeds is of order $10^{-3}$ for both the raw and averaged density models. The anisotropic components in the initial conditions add a statistically significant amount of information, increasing the MI by a factor of 1.5.

Despite being statistically significant, this increase in MI does not correspond to a useful qualitative improvement in predicting halo mass.
To show this, in Fig.~\ref{fig:raw_vs_averaged_MI}, the grey horizontal lines indicate the value of the MI for fictitious halo mass `models' that were constructed by adding Gaussian noise to the ground truth values with standard deviations of 1.5, 1.0, 0.75, 0.5 dex, respectively.
By comparing the increase in MI due to the anisotropic components in the initial density field with these fictitious benchmark predictions, we see that while the scatter in predictive accuracy decreases by $\sim 0.2$ dex, the overall scatter still remains high at $\sim 0.7$ dex. These results confirm that while we have quantified a statistically significant amount of information contained in the anisotropic aspects of the initial density field about the final halo mass, it does not constitute sufficient information to provide a qualitative improvement in the initial conditions-to-halo mass mapping.

\begin{figure*}
\centering
\includegraphics[width=\textwidth]{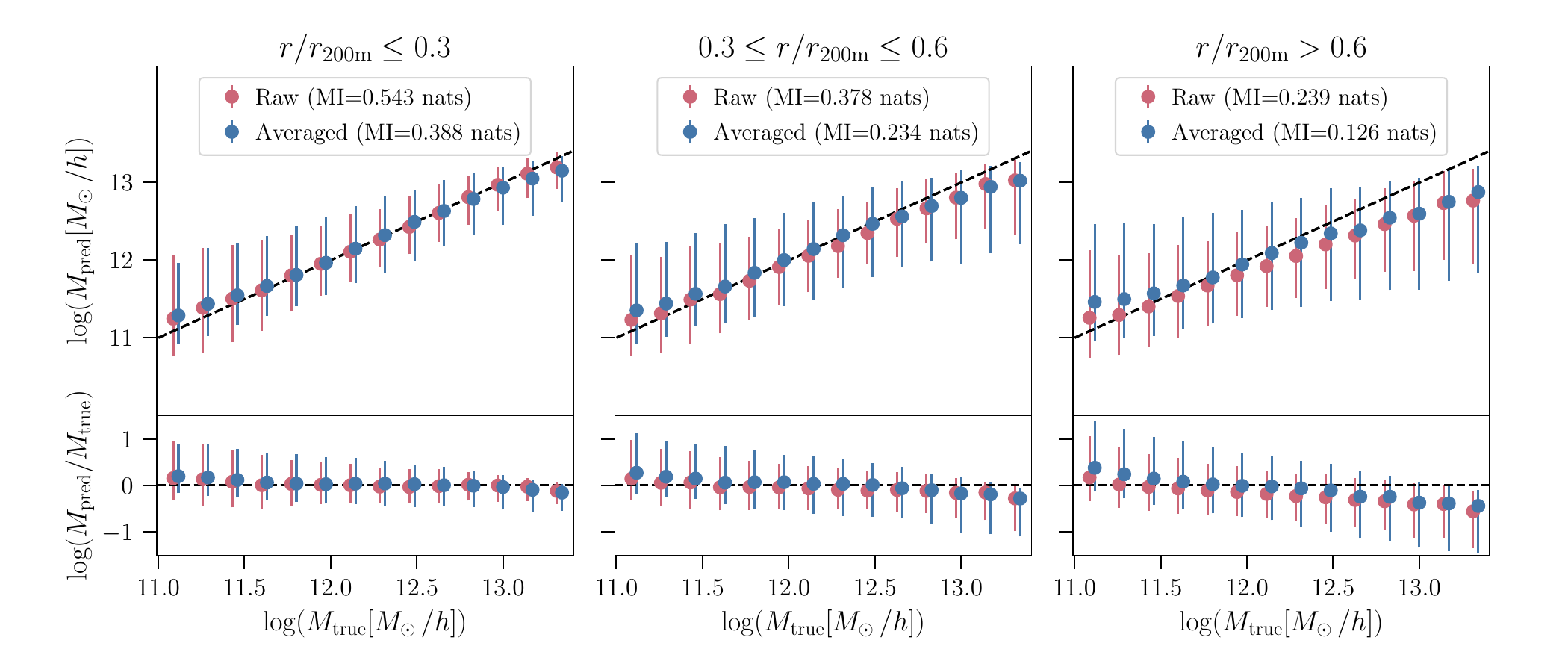}
\caption{Halo mass predictions for particles that reside in different locations inside the halos: those located in the inner region of the halo ($r \leq 0.3 \, r_{\rm 200m}$; left panel), those in a intermediate region ($0.3 < r/r_{\rm 200m} \leq 0.6$); middle panel), and those in the outskirts of halos ($r > 0.6 \, r_{\rm 200m}$; right panel). The MI between predicted and ground truth halo mass values is indicated in the legend box of each panel. While the change in MI between raw-density and averaged-density models is statistically significant, the decrease in the scatter of the predictions is not sufficient to provide a qualitative improvement in the initial conditions-to-halo mass mapping.}
\label{fig:raw_vs_averaged_rbins}
\end{figure*}

\subsection{Dependence on particles' radial position}

We next compare the accuracy of the predictions across particles that reside in different locations inside the halos. Fig.~\ref{fig:raw_vs_averaged_rbins} shows the predictive accuracy of the models for three sets of particles that were split by their location inside the halos. Particles are split into those that live in the innermost region of halos (left panel) i.e., $r \leq 0.3 \, r_{\rm 200m}$, where $r$ is the radius of the particle from the centre of its host halos and $r_{\rm 200m}$ is the halo virial radius, those in a mid-region (middle panel) i.e., $0.3 < r/r_{\rm 200m} \leq 0.6$ and those in the outskirts of halos (right panel) i.e., $r > 0.6 \, r_{\rm 200m}$.
We find that the particles with the most accurate predictions are those in the innermost regions of the highest-mass halos. This is true for both the averaged-density and raw-density models. Particles that live in the outskirts of high-mass halos have larger bias and variance in their predictions: this is partially due to the fact that sometimes the input sub-box volume does not include the full extent of the region that will later collapse into a high-mass halo. Therefore, when the input sub-box is centered on an outskirt particle, its volume does not cover a large fraction of the proto-halo region, making it difficult for the CNN to infer the correct final halo mass. This explains why outskirts particles tend to yield larger errors than inner particles. For halos of smaller mass, the distribution of predictions for particles in different locations inside the halos share similar variances. One way to improve the predictions of outskirt particles in high-mass halos would be to use larger sub-box volumes as inputs. However, adopting a larger sub-box at the same resolution was not possible due to computational limitations in memory consumption on one Tesla V100 GPU. Similarly, decreasing the resolution to accommodate for a larger volume would cause the predictions to worsen for particles in low-mass halos. Despite this, the performance of the CNN remains similar to or better than the predictive accuracy of analytic models (see Appendix \ref{sec:analytic}).

In all panels, the MI values indicated in the legend box show a statistically significant gain in information when adding anisotropic aspects  to the inputs; however, this again translates into a qualitatively small reduction in the scatter of the predictions. Therefore, our conclusion that the anisotropic information in  the inputs does not yield a useful improvement in halo mass modelling is valid for all particles, regardless of their location inside the halo.

\section{Demonstrating the ability of the CNN to extract features}
\label{sec:z0}
One fundamental assumption behind this interpretation of our results is that the CNN is capable of capturing features across the range of different scales present in the input sub-box. If instead the CNN's ability to learn were limited, then we could not exclude the possibility that there exists additional information in the inputs which affects halo collapse, but which the CNN is unable to learn. This motivated us to perform tests showing that the same CNN model can return highly accurate predictions, when presented with the information to do so. In particular, we wanted to create a test as closely related to the real initial conditions-to-halo mass problem, which would specifically demonstrate the ability of the CNN to extract features from the density field, on all scales probed by the input sub-boxes, and return halo mass predictions that are consistent with expectations.

To do this, we tested the performance of the model in a scenario where we could compare the predictions of the CNN to our expectations. We trained the CNN to learn the mapping between the non-linear density field at the present time ($z=0$) and the mass of the resulting halos. This mapping is effectively given by an algorithm which first identifies the boundary of a halo based on a fixed density threshold, similar to the friends-of-friends algorithm used to identify halos in the simulation, and then computes the mass enclosed within such halo. To do this, the CNN must be able to simultaneously extract features at a number of different scales; from that of the boundary of the lowest mass halos up to that of the most massive ones. As this is a more straightforward mapping than that between the initial conditions density field and the final halo masses, we expect the CNN to return near-perfect predictions.

\begin{figure*}
\centering
	\includegraphics[width=\textwidth]{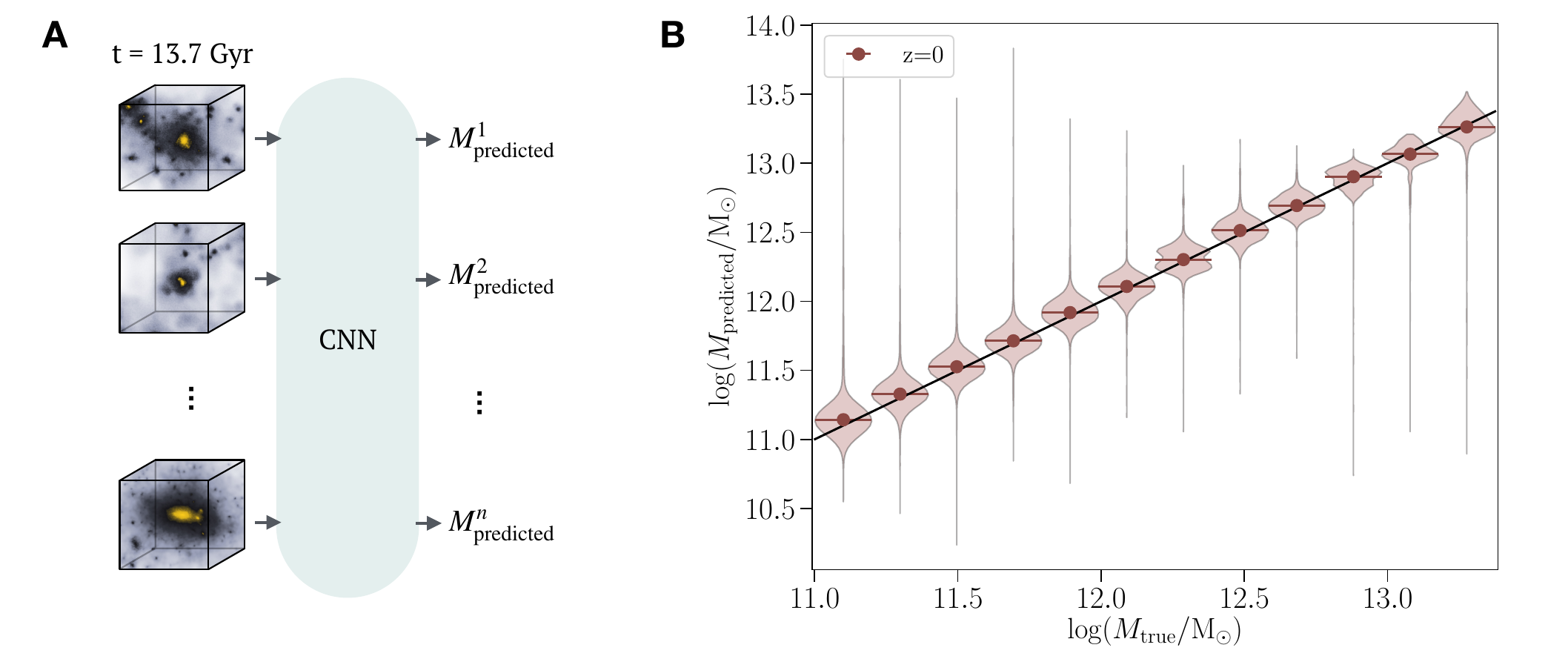}
    \caption{Halo mass predictions returned by a CNN trained on the non-linear density field at $z=0$. The predictions are shown in the form of violin plots i.e., distributions (and their medians) of predicted halo masses of particles within evenly-spaced bins of true logarithmic halo mass. The predictions are in excellent agreement with their respective ground truth halo masses, yielding a correlation coefficient $r=0.96$.}
    \label{fig:z0_violins}
\end{figure*}

Similar to the $z=99$ case, we provided the CNN with the non-linear density field in cubic sub-regions of the simulation, centered at each particle's position. The inputs are given by the non-linear density field $\delta \left( \mathbf{x}, t_\mathrm{final} \right)$ at $z=0$ in a sub-volume centered at each particle's position. As for the $z=99$ case, the density is estimated based on the particles' positions. We revisited our choices of box size and resolution of the 3D sub-box, as the scales of interest at $z=0$ naturally differ from those in the initial conditions. We fixed the resolution to that used for the $z=99$ case, $N=75^3$, and chose a box size of $L=1.5 \, \mathrm{Mpc} \, h^{-1}$, which approximately corresponds to the virial radius of a halo with mass $M=10^{14}\, \mathrm{M}_\odot$. These choices resulted in a voxel length $l_\mathrm{voxel} \sim 30 \, \mathrm{kpc} \, h^{-1}$, which is approximately equivalent to half the virial radius of a $M=10^{10}\, \mathrm{M}_\odot$ halo. Given that the box captures the virial radius of the largest and smallest halos probed by our simulations, we expect the input boxes to contain the information required by the algorithm to learn the density field-to-halos mapping. To summarize, the $z=99$ and $z=0$ settings use as inputs the density fields at $t=17.1$ Myr and $t=13.7$ Gyr after the Big Bang respectively, while keeping all other choices about the model's architecture and its hyperparameters identical. 

Fig.~\ref{fig:z0_violins} shows the predictions of the CNN when trained on the $z=0$ non-linear density field. Just like in Fig.~\ref{fig:CNN_z99}B, the predictions are illustrated in the form of violin plots, showing the distributions of predicted halo masses in bins of true mass. As expected, the predictions show good agreement with the true halo mass labels, yielding a correlation coefficient $r=0.96$, where $r=1$ implies an exact linear relationship. The presence of a very low number of outliers, that make up the visible tails of the violin plots, is expected from any machine learning model trained from a finite dataset; the fraction of particles with predicted mass outside the 3$\sigma$ interval of $\log(M_{\rm predicted}/M_{\rm true})$ is only 0.4\%. Since the predictions are highly accurate throughout the full mass range of halos, the CNN must be able to identify the relevant features from the density field on all scales within the input sub-box and yield predictions within expected accuracy. 

Crucially, the volume and resolution of the input sub-boxes were adapted to resolve the smallest and largest scales of interest at $z=0$; therefore, this test confirms the ability of the CNN to extract features from the smallest to the largest accessible scales of the inputs. Consequently, we expect the same architecture to also have the capability to extract features on multiple scales within the $z=99$ linear density field. However, since the features in the initial conditions have a much more complex relationship with halo mass, the predictions will naturally be less accurate than the $z=0$ case. 

\section{Discussion and conclusions}
\label{sec:conclusions}
We have presented a deep learning framework, capable of learning final halo masses directly from the linear density field in the initial conditions of an $N$-body simulation. The overall goal of our work is to learn about physical aspects of the early Universe which impact the formation of late-time halos using the results of deep learning, without the need to featurize the initial conditions. To do this, we require a deep learning framework that allows for the interpretability of its learning; for example, in understanding the features assembled by the convolutional layers and how these map onto the final predictions. In this work, we removed part of the information from the inputs and re-trained the CNN to test the impact of this on the accuracy of the final predictions. 

This allowed us to quantify in full generality the amounts of information in the isotropic and anisotropic components of the initial density field about final halo masses. We found a small, albeit statistically-significant amount of additional information in the anisotropic component over that contained in the isotropic component of the initial density field; however, this corresponds to only a 0.2 dex decrease in a scatter of 0.9 dex in predictive accuracy. Thus, the addition of anisotropic information does not yield qualitative improvements in the initial conditions-to-halo mass mapping in the range $\log \left( M/\mathrm{M}_{\odot} \right) \in \left[ 11, 13.4 \right]$.
In practice, the information in the initial conditions gleaned by the deep learning model to infer halo masses is equivalent to that captured by spherical averages over the initial density field. Our conclusions do not change if we train on the initial potential field instead of the density field, demonstrating that long-range gravitational effects do not significantly affect the local process of halo collapse (see Appendix \ref{sec:potential}). A crucial test of robustness of our framework was to demonstrate that the deep learning model can effectively extract spatially-local features on all scales probed by the input sub-boxes and yield robust halo mass predictions that match expectations for a simpler test-case scenario.

The idea of removing or changing parts of the data and retraining has previously been used in the deep learning community as part of a data engineering step. For example, adversarial examples are visually imperceptible perturbations added to the data that yield catastrophic failures in the CNN predictions \cite{Szegedy2013}; these are often used to test the robustness of CNNs. In the context of astrophysics, inputs are often modified to include different levels of noise in simulated data \cite{Fluri2018,Schmelzle2017}; this is also done to test the robustness of the model against noise. In our work, modifications to the input data are made to remove specific physical aspects from the initial density field; the aim is to verify whether the removed information plays a role in inferring the final output. To our knowledge, this is the first time these techniques have been adopted for generating a physical interpretation of a CNN model within a cosmological setting.

Our results lead to a re-evaluation of the current understanding of gravitational collapse based on more traditional analytic and semi-analytic approaches. Existing studies based on the ellipsoidal collapse model \cite{ShethMoTormen2001, ShethTormen2002, Monaco2002, castorina2016, Paranjape2013} incorporate tidal shear effects either indirectly or in the form of free parameters calibrated to numerical simulations. Peak-patch theories \cite{Bond&Myers1996, Bond&Myers19962, Stein2018} do not quantify the impact of the added tidal shear information on the predictive power for halo mass compared to spherically-averaged density alone. Therefore, the models do not in themselves demonstrate a significant role for tidal shear in determining final halo mass. Our work focuses on providing a direct test of the importance of anisotropic information, and not just tidal shear, and shows that this does not in fact play a significant role in establishing final halo masses in the range $\log \left( M/\mathrm{M}_{\odot} \right) \in \left[ 11, 13.4 \right]$. Consequently, building better theoretical models of halo collapse requires incorporating information beyond the initial conditions, such as approximations to the gravitational evolution of the density field. Our results illustrate the promise of deep learning frameworks as powerful tools for extracting new insights into cosmological structure formation.

\section*{Data and code availability}
The software used to generate the simulations are available at \href{https://github.com/pynbody/genetIC}{https://github.com/pynbody/genetIC} to generate the initial conditions, and at \href{https://wwwmpa.mpa-garching.mpg.de/gadget/}{https://wwwmpa.mpa-garching.mpg.de/gadget/} to run the $N$-body simulations. The parameter files for generating the initial conditions and simulations can be made available from the authors upon reasonable request. The data used in this work, including simulations and training/validation/test sets, can be freely downloaded from \href{https://console.cloud.google.com/storage/browser/deep-halos-data;tab=objects?forceOnBucketsSortingFiltering=false&project=deephalos&prefix=&forceOnObjectsSortingFiltering=false}{Google Cloud Storage}. The code used to conduct the analysis is publicly available at \href{https://github.com/lluciesmith/DeepHalos}{github.com/lluciesmith/DeepHalos}.

\section*{Author Contributions}
\textbf{L.L.-S.}: led the project; project conceptualisation; methodology; software; obtained, validated, and interpreted results; writing - original draft, editing, final; funding acquisition. \textbf{H.V.P. \& A.P.}: project conceptualisation; methodology; investigation, validation \& interpretation; supervision; writing - editing; funding acquisition. \textbf{B.N.}: methodology; investigation; writing - editing; provided resources. \textbf{J.T.}: writing - review; provided resources.

\begin{acknowledgments}
We thank Davide Piras for assistance with GMM-MI. LLS thanks Corentin Cadiou, Joao Caldeira, Andres Felipe Alba Hernandez, Eiichiro Komatsu, Michael Maire, Manuel Valentin and Samuel Witte for useful discussions. LLS thanks Stephen Stopyra for help in generating the simulations. HVP thanks Justin Alsing and Daniel Mortlock for useful discussions. LLS acknowledges the hospitality of the Fermi National Accelerator Laboratory, where part of this work was completed, and thanks Josh Frieman and Brian Nord for kindly hosting. HVP acknowledges the hospitality of the Aspen Center for Physics, which is supported by National Science Foundation grant PHY1607611. This project has received funding from the European Union’s Horizon 2020 research and innovation programme under grant agreement No. 818085 GMGalaxies. LLS was also supported by the Science and Technology Facilities Council. HVP was partially supported by the  research  project  grant  ``Fundamental  Physics  from  Cosmological  Surveys'' funded by the Swedish Research Council (VR) under Dnr 2017-04212. This project has received support from the research project grant ``Understanding the Dynamic Universe'' funded by the Knut and Alice Wallenberg Foundation under Dnr KAW 2018.0067. AP was supported by the Royal Society. This work was partially supported by the Visiting Scholars Award Program of the Universities Research Association, and by funding from the UCL Cosmoparticle Initiative. This work used computing facilities provided by the UCL Cosmoparticle Initiative; and we thank the HPC systems manager Edd Edmondson for his committed support. This work was also partially supported by Wave 1 of The UKRI Strategic Priorities Fund under the EPSRC Grant EP/T001569/1, particularly the ``AI for Science'' theme within that grant \& The Alan Turing Institute, and by the Benchmarking for AI for Science at Exascale (BASE) project under the EPSRC Grant EP/V001310/1. Work supported by the Fermi National Accelerator Laboratory, managed and operated by Fermi Research Alliance, LLC under Contract No. DE-AC02-07CH11359 with the U.S. Department of Energy. The U.S. Government retains and the publisher, by accepting the article for publication, acknowledges that the U.S. Government retains a non-exclusive, paid-up, irrevocable, world-wide license to publish or reproduce the published form of this manuscript, or allow others to do so, for U.S. Government purposes.
\end{acknowledgments}

\appendix
\section{The gravitational potential vs the density field as input}\label{sec:potential}

\begin{figure}
	\centering
         \includegraphics[width=\columnwidth]{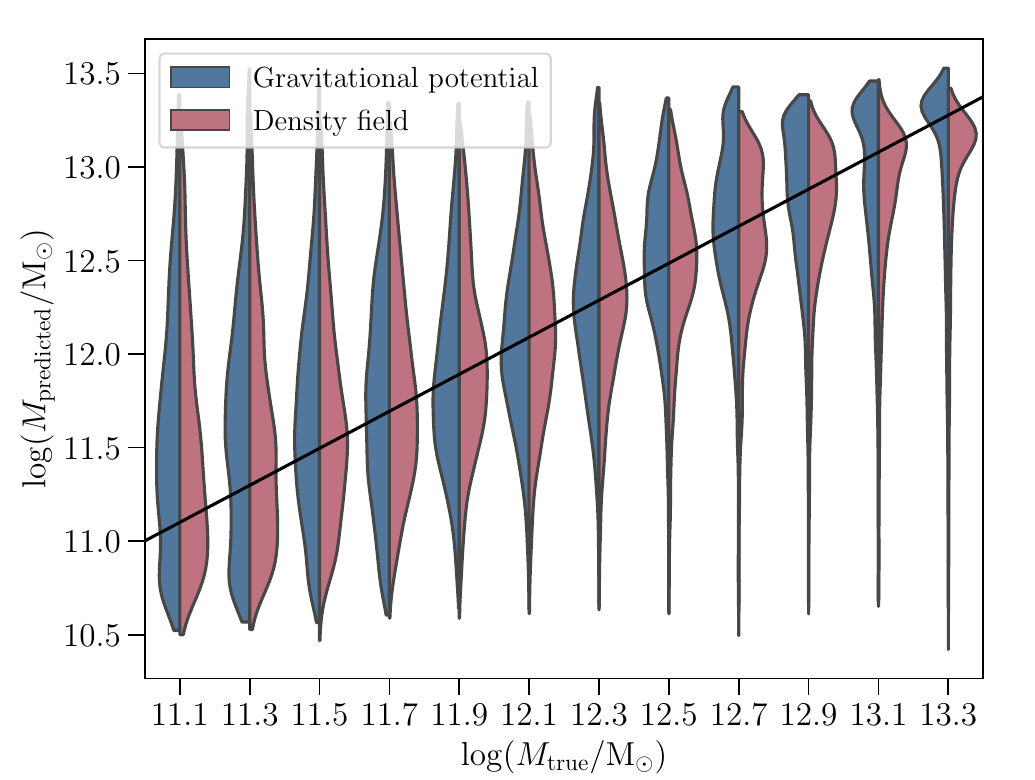}
    \caption{We compare the predictions returned by the CNN when trained on the initial density field and when trained on the gravitational potential field. The two models yield consistent predictions, meaning that long-range gravitational effects do not impact the final mass of halos and that any information needed about the potential can be retrieved from the density field by the CNN.}
    \label{fig:potential}
\end{figure}

The initial density field contains all the information to fully describe the initial conditions of the Universe. Other fields, such as the potential field, its gradient or its Hessian, can be derived directly from the density field in the whole simulations box via the Poisson equation. The solution to the Poisson equation is a convolution, meaning that information about the potential field is accessible to the CNN from the density field. However, the one-to-one correspondence between potential and density fields is only strictly valid in the whole simulation box. Since our inputs are limited to sub-regions of the simulations, the information contained within the density field is not \textit{exactly} the same as that carried by the potential field within that same sub-region. In particular, the density field within a sub-volume of the simulation excludes information about long-range gravitational effects which would instead be accessible through the potential field.

To test whether long-range gravitational effects contain relevant information for the CNN to predict final halo masses, we trained the CNN on the gravitational potential field instead of the density field. We computed the potential field across the entire simulation by solving the Poisson equation, and replaced the initial density field within each input sub-box with the gravitational potential field within that same region. We re-trained the CNN model using this new set of inputs and compared the predictions returned by the CNN when trained on the density field and when trained on the gravitational potential field. Fig.~\ref{fig:potential} compares the distributions of predicted halo masses within bins of true halo masses for the two cases. Each distribution is estimated using a kernel density estimate over the set of discrete mass values predicted by the model. We found that the predictions from the two models are consistent, despite small variations in the predicted distributions at the high-mass end. 
We quantified the significance of these differences by computing the MI (Eq.~\eqref{eq:MI}) between predicted and ground truth halo mass values, for both the potential and density models. The former yields $I(\log M_\mathrm{pred}, \log M_\mathrm{true}) = 0.399 \pm 0.004$, which is similar to the MI of the density model $I(\log M_\mathrm{pred}, \log M_\mathrm{true}) = 0.370 \pm 0.004$. Therefore, we conclude that small variations in the predictions from the two models present at the high-mass end do not significantly impact the overall predictivity of the model.

This test demonstrates that long-range gravitational effects do not impact the final mass of halos in any significant way. Any information needed about the potential can be retrieved from the density field by the CNN, despite the fact that this information is expected to reside on larger scales. We therefore expect our inputs -- the density field within sub-volumes of the simulation -- to capture all relevant information in the initial conditions about final halo mass.

\section{CNN architecture}
\label{sec:architecture}
Our deep learning architecture consists of six convolutional layers, all but the first one followed by max-pooling layers, and three fully-connected layers. The convolutions were performed with 32, 32, 64, 128, 128, 128 kernels for the six convolutional layers, respectively -- all with a stride of 1 and zero-padding. The initial weights of the kernels in a layer were set following the Xavier initialization technique \cite{glorot10}, which randomly draws values from a uniform distribution bounded between $\pm \sqrt{6/\left( n_i + n_{i+1}\right) }$, where $n_i$ and $n_{i+1}$ are the number of incoming and outgoing network connections to that layer. The kernels have size $3\times 3 \times 3$ in all convolutional layers, meaning that the first layer learns features on scales of $0.6 \, \mathrm{Mpc} \, h^{-1}$. As more convolutional layers are stacked on top of each other, the algorithm becomes sensitive to features at increasing scales. In this way, both local and global information are able to propagate through the network. We applied a a non-linear activation function to every feature map given by a leaky rectified linear unit (LeakyReLU) \cite{Nair2010}
\begin{equation}
	f(x) = \begin{cases}  
	x &\text{ for } x \geq 0, \\ 
	\beta \times x &\text{ for } x < 0,
	\end{cases}
	\label{eq:relu}
\end{equation}
with $\beta = 0.03$. A LeakyReLU activation, with $\beta$ of order $10^{-2}$, is a common choice that has proved successful in many deep learning applications \cite{Nwankpa2018}. The feature maps are then fed to max-pooling layers, which reduce their dimensionality by taking the average over $2\times2\times2$ non-overlapping regions of the feature maps. The CNN is inherently translationally invariant due to the combination of convolutional and pooling layers in the architecture; we do not incorporate any further symmetries in the network.

After the sixth convolutional layer and subsequent pooling layer, the output is flattened into a one-dimensional vector and fed to a series of 3 fully-connected layers, each made of 256 and 128 and 1 neuron, respectively. The non-linear activation function of the first two layers is the same ReLU activation (\eqref{eq:relu}) as that used in the convolutional layers, whereas the last layer has a linear activation in order for the output to represent halo mass. The weights were initialized using the same Xavier initialization technique used for the kernel weights of the convolutional layers. Regularization in the convolutional and fully-connected layers was incorporated in the form of priors over the parameters of the model, as explained in the next subsection. 

We chose the architecture that returned the best performance (i.e., the lowest loss score on the validation set after convergence) amongst many, but not all, alternative models with different choices of architecture-specific and layer-specific hyperparameters. We investigated the change in the validation loss in response to the following modifications: adding batch-normalization layers; introducing dropout; varying the amount of dropout; adding/removing convolutional layers and/or fully-connected layers; increasing/decreasing the number of kernels/neurons in each convolutional/fully-connected layer; changing the weight initializer; changing the convolutional kernel size.  In all cases, we found that the final loss score either increased or showed no change compared to that of the architecture retained in this work. We leave further hyperparameter exploration, including changes to the optimizer, the addition of skip connections, and other variations in the architecture, to future work.

\section{The loss function}
\label{sec:loss}
A neural network can be viewed as a probabilistic model $p \left( \bol{y} \mid \bol{x}, \bol{w} \right)$, where given an input $\x$, a neural network assigns a probability to each possible output $y$, using the set of parameters $\bol{w}$. The parameters are learnt via maximum likelihood estimation (MLE): given a set of training examples $\mathcal{D} = \{x_i, d_i \}_{i=1}^{N}$, the optimal weights are those that minimize the negative log-likelihood, $\ln \left[ p \left( \mathcal{D} \mid \bol{w} \right) \right] = \sum_{i=1}^N \ln \left[ p \left(  d_i \mid x_i, \bol{w} \right) \right]$, more generally called the loss function $\Loss$ in the machine learning community. The issue is that deep neural networks are generally over-parametrized; it has in fact been demonstrated that there exists a major redundancy in the parameters used by a deep neural network \cite{Denil2013}. This means that when minimizing the negative log-likelihood with a deep neural network model, one almost always encounters the problem of overfitting. The algorithm tends to fit the samples of the training data $\mathcal{D}$ extremely well but fails to learn patterns that are generalizable to unseen data. To overcome this issue, one modifies the loss function of the neural network in such a way that prevents the algorithm from overfitting and improves its generalizability. This is known as regularization.

We introduce regularization by adopting priors over the weights. Following Bayes' theorem, the goal of the neural network then becomes to maximize the posterior distribution $p \left( \bol{w} \mid  \mathcal{D} \right) = p \left( \mathcal{D} \mid  \bol{w} \right) p \left( \bol{w} \right)$, rather than the likelihood $p \left( \mathcal{D} \mid  \bol{w} \right)$. The loss function, $\Loss$,  is then given by
\begin{equation}
\Loss = - \ln \left[ p(\bol{w} \mid \mathcal{D}) \right] = - \ln \left[ p \left( \mathcal{D} \mid  \bol{w} \right) \right]  - \ln \left[ p(\bol{w}) \right],
\label{eq:posterior_weights}
\end{equation}
where the first is the likelihood term, or predictive term $\Loss_\mathrm{pred}$, and the second is the prior term, or regularization term $\Loss_\mathrm{reg}$, as in \eqref{eq:loss_overview}. If $\bol{w}$ are given a Gaussian prior, this yields L2 regularization; if $\bol{w}$ are given a Laplacian prior, then one obtains L1 regularization. The advantage of this form of regularization is that it can be incorporated in terms of priors on the weights. There exist many other regularization techniques, including for example dropout, but we choose to focus on those that have a direct Bayesian interpretation.

Technically, the parameters optimized during training include not just the weights, but also the biases. These consist of a constant value that is added to the product of inputs and weights for every kernel (neuron) in a convolutional (fully-connected) layer. These parameters add little flexibility to the model and are therefore typically not responsible for overfitting. Therefore, we choose not to consider setting priors on the biases as they do not require regularization.

\subsection{The choice of the likelihood function}
\begin{figure*}
\centering
	\includegraphics[width=0.9\textwidth]{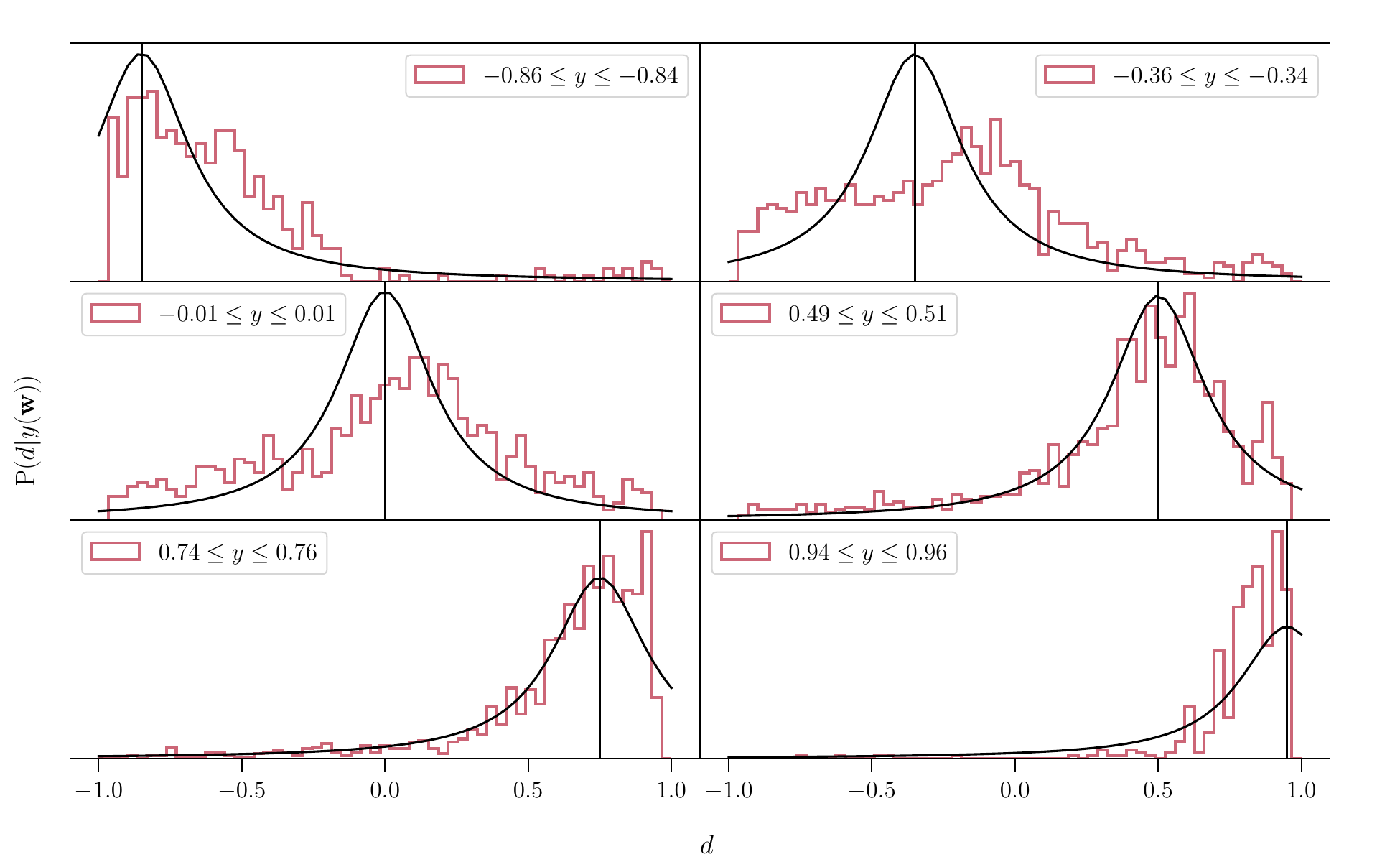}
    \caption{Distribution of $d$ for fixed slices in $y$, where $d$ is the ground truth logarithmic halo mass variable rescaled to the range $[ -1, 1 ]$ and $y$ is the predicted value returned by the CNN in rescaled units. The black line is the likelihood function given the value of $\gamma$ at the epoch where the validation loss reaches its minimum, $\gamma=0.26$.}
    \label{fig:likelihood_fit}
\end{figure*}

We denote $d = d \left( \x \right)$ as the ground truth variable rescaled to $\left[-1, 1\right]$ and $y = y \left( \x, \bol{w} \right)$ as the prediction returned by the CNN model with weights $\bol{w}$, for a given set of rescaled inputs $\bol{x}$. The likelihood function describes the distribution of ground truth values $d$ for a given value of predicted output $y$, returned from the neural network model with weights $\bol{w}$. A typical choice in the field is that of a Gaussian or Laplacian likelihood, yielding the popular mean-squared-error or mean-absolute-error losses. For our problem, we found that a Gaussian distribution is a poor description of the training data: the distributions of $d$ for fixed values of $y$ contain long tails, especially when $y$ is close to the boundaries $y=-1$ and $y=1$, that a Gaussian distribution fails to account for. Not accounting for these tails led to biased predictions, especially towards the boundaries. Instead, we chose a Cauchy distribution function for the likelihood, characterized by the scale parameter $\gamma$, which has broader tails than a Gaussian and a well-defined form for its conditional distribution function.

The negative log-likelihood then becomes
\begin{align}
\begin{split}
- \ln &\left[ p \left(d \mid y, S \right) \right] = \frac{1}{N} \sum_{i=1}^N\left[\ln(\gamma) + \ln \left[ 1 + \left( \frac{d_i - y_i}{\gamma}\right)^2 \right] \right. \\
 &+ \left. \ln \left[\arctan\left( \frac{d_\mathrm{max} - y_i}{\gamma} \right) - \arctan \left( \frac{d_\mathrm{min} - y_i}{\gamma} \right) \right] \right]
 \label{eq:lik_Cauchy_selection}
 \end{split}
\end{align}
for a Cauchy likelihood function with scale parameter $\gamma$, under a top-hat selection function $S$ over the ground truth variable, $p \left( S \mid d \right) = \Theta \left( d_\mathrm{max} - d \right) \Theta \left( d - d_\mathrm{min} \right)$, where $\Theta$ is the Heaviside step function. The latter arises from the fact that, by construction, the rescaled ground truth is restricted to $d_\mathrm{min} \leq d \leq d_\mathrm{max}$, where $d_\mathrm{min} = -1$ and $d_\mathrm{max}=1$. This selection function was needed in order to correctly model the loss at the boundaries. The first two terms in \eqref{eq:lik_Cauchy_selection} arise from the Cauchy likelihood: the first is effectively a prior on $\gamma$ which is insensitive to the predictions $y$, and the second measures the difference between predicted and ground truth values weighed by the scale parameter $\gamma$. The third term in  \eqref{eq:lik_Cauchy_selection} comes from accounting for the selection function $S$. The normalization factor $1/N$ is not part of the negative log-likelihood but is typically introduced in the loss function so that the loss is insensitive to the size of the training set (or, of the batch size if performing batch gradient descent when training). The scale parameter $\gamma$ determines the half-width at half-maximum of the Cauchy distribution. Since the optimal value of $\gamma$ is not known a priori, we optimize that parameter during training using back-propagation, together with the rest of the weights and biases optimized by the network. To test the robustness of simultaneously optimizing loss function hyperparameters and network weights, we retrained the network with $\gamma$ fixed to its best-fit value and found no significant change in the performance of the network. The above likelihood term in the loss function satisfies our desiderata of having a heavy-tailed probability distribution function and accounting for the restricted range of ground truth $d$.

The expression in \eqref{eq:lik_Cauchy_selection} is valid under the condition that $d, y \in \left[-1, 1\right]$. However, since the activation function in the last layer is given by the unbounded linear function $\sigma(z)=z$, the predictions can technically take any value $y \in \mathbb{R}$. To solve this, we introduced a super-exponential function, denoted as $f(y)$, in the regime $|y| \geq 1$, to counter balance the Cauchy limits at the boundaries and sharply disfavour predictions outside the interval $\left[ -1, 1\right]$. The function is continuously matched to the Cauchy distribution at the boundaries $y = \pm 1$. The likelihood term of the loss function $\Loss_\mathrm{pred}$ is then given by a piecewise function conditional on $y$;

\begin{align}
\begin{split}
\Loss_\mathrm{pred} =  \frac{1}{N} \sum_{i=1}^N \left[  \right. &  \left. - \ln p \left(d_i \mid y_i, S\right) \Theta \left( |y| + 1 \right) \right. \\
& \left. + f(y_i)  \Theta\left( |y| - 1 \right) \right].
\end{split}
\label{eq:likpiecewise}
\end{align}

Fig.~\ref{fig:likelihood_fit} compares the form of the likelihood, given the optimized value for $\gamma$ returned by the model, compared to the empirical distribution of ground truth values at fixed slices in $y$, the predicted variable returned by the trained CNN, for the training set samples. The Cauchy likelihood provides a good fit to the empirical likelihood distribution of the model. However, we note that for small values of $y$, the fit could have been improved by adopting a two-tailed Cauchy likelihood function. Further flexibility could be provided by using a Student's t-distribution. We leave this to future work.

\subsection{The choice of priors: regularization and model compression}
We adopt weight priors that can simultaneously (i) improve the optimization during training by preventing overfitting and (ii) compress the neural network model into the least number of parameters without loss in performance. Regularization and model compression are very much related: these tasks can be achieved simultaneously by minimizing a properly defined cost function. We selected weight priors that penalize large values (for regularization) and induce sparsity (for model compression). To regularize the network, we adopted weight priors that promote smaller values, as these typically lead to more generalizable solutions. We chose Gaussian priors for the weights of the convolutional layers and Laplacian priors for the weights of the fully-connected layers, which penalize the sum of the squared values or the sum of the absolute values of the weights, respectively. The choice of Laplacian prior has the additional benefit that it induces sparsity on the weights by driving most weights to be zero; a Laplacian prior thus combines the idea of model compression and regularization. For model compression, our aim is to induce a more compact network with the smallest number of non-zero neurons in the fully-connected layers. To do this, we adopted the group Lasso formulation \cite{Scardapane2017} which imposes group-level sparsity, meaning that all the variables in that group are either simultaneously set to 0, or none of them are. For the case of fully-connected layers, a group is equivalent to an entire neuron.

The log priors over the weights become
\begin{align}
\begin{split}
\ln \p(\bol{w}) = \alpha \left[\sum_{l \in l_c} \sum_{p=1}^{P_l} \left( w^{(l)}_p \right)^2 \right. & \left. + \sum_{l \in l_d} \sum_{q=1}^{Q_l} |w^{(l)}_q|  \right. \\
& \left. + \sum_{l \in l_d} \sum_{i=1}^{N_{l-1}}  \left[ \sum_{j=1}^{N_l} w_{ij}^2 \right]^{1/2} \right],
\end{split}
\label{eq:priors}
\end{align}
where the first term is a Gaussian prior over each of the $P_l$ weights of each convolutional layer $l_c$, $w^{(l)}_p$, the second term is a Laplacian prior over each of the $Q_l$ weights of each fully-connected layer $l_d$, $w^{(l)}_q$, and the third term is a group Lasso prior over the set of weights that determine the connections between a single neuron in the $(l-1)$-th layer and all the neurons in the $l$-th layer. The idea of group-level sparsity can also be applied to convolutional layers, where a single group is given by the collection of weights from a single kernel of the layer. This can be thought of as a feature selection method, in that it removes entire kernels (and thus the feature represented by that kernel) within each convolutional layer. Given that our network is relatively small, we chose not to perform feature selection; we leave this for future work.

The prior term $\ln \left[ \p(\bol{w}) \right]$ is added to the likelihood term in the loss function, as in \eqref{eq:posterior_weights}. The regularization parameter $\alpha$ in \eqref{eq:priors} weighs the prior term relative to the likelihood term in the loss function. Its value sets the balance between an overly-complex model (which overfits and has a high variance) and an overly-simple model (which underfits and has a high bias). We optimized this parameter, in combination with the learning rate, using cross-validation.

\section{Training and optimization}
\begin{figure}[t]
\centering
	\includegraphics[width=\columnwidth]{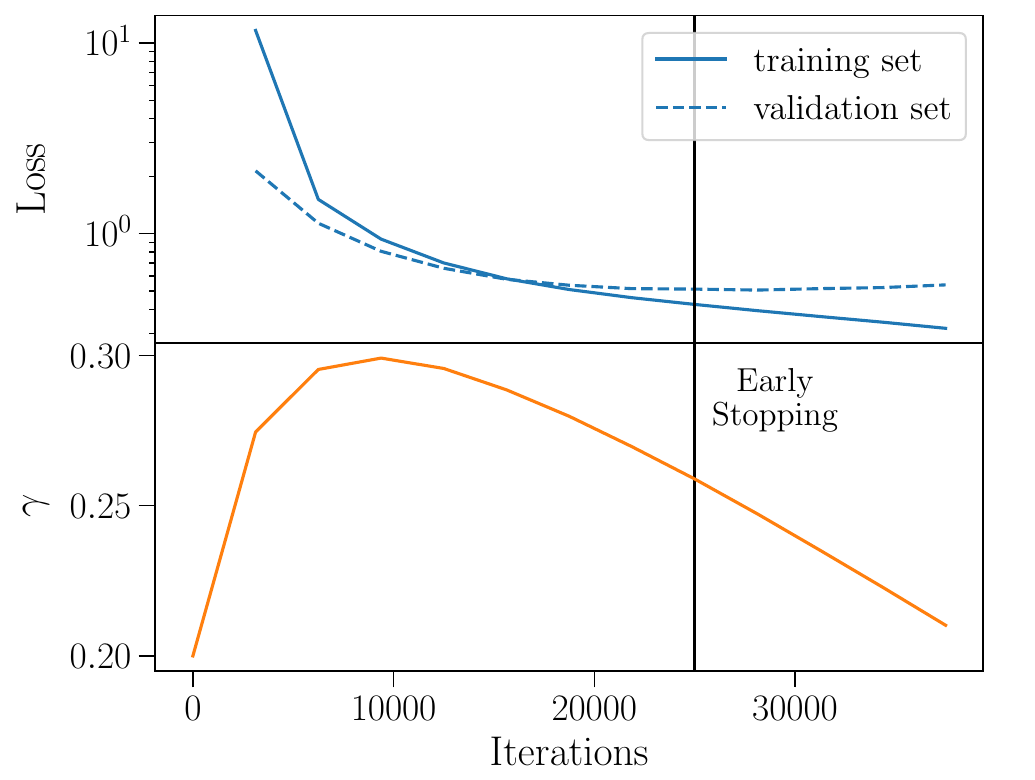}
    \caption{\textit{Top panel:} The loss function evaluated for the training set and the validation set at each batch iteration. Early stopping was employed to interrupt the training at the epoch where the validation loss reaches its minimum. The weights of the CNN at the early-stopping iteration were retained. \textit{Bottom panel:} The half-width at half-maximum parameter of the Cauchy likelihood function, $\gamma$, as a function of iteration.}
    \label{fig:loss_curve}
\end{figure}

The algorithm was trained on $200$,$000$ particles, randomly drawn from the ensemble of particles of $20$ simulations based on different realizations of the initial conditions. We validated the model using 10,000 particles from a single simulation, and tested it on 99,950 particles from four independent simulations. We did not perform data augmentation to increase the size of the training set since we had available a large number of training samples and therefore opted for testing the impact of adding new (independent) samples instead.
We found no improvement in the performance of the algorithm as we added to the training set an additional $300$,$000$ particles from another independent simulation, implying that our choices were sufficient to yield a training set representative of the mapping between initial conditions and halos.
We further investigated changes to the training set, such as re-balancing the training set to have the same number of particles in low- and high-mass halos, and found no change in the accuracy of the predictions. The training set was sub-divided into batches, each made of $64$ particles. Batches were fed to the network one at a time, and each time the CNN updates its parameters according to the samples in that batch.

Training was done using the \texttt{AMSGrad} optimizer \cite{Reddi2019}, a variant of the widely-used \texttt{Adam} optimizer \cite{Kingma2014}, with a learning rate of 0.00005. The learning rate was optimized via cross-validation, together with $\alpha$, the parameter weighting the regularization term in the loss function. The number of trained parameters in the network is $2,108,258$. Fig. \ref{fig:loss_curve} shows the loss function (\textit{upper panel}) evaluated for the training and validation sets, and the value of the parameter $\gamma$ in the likelihood term of the loss (\textit{lower panel}) as a function of the number of iterations.  Early stopping was employed to interrupt the training at the epoch where the validation loss reaches its minimum value; the early-stopping iteration is shown as a vertical grey line in Fig.~\ref{fig:loss_curve}. The final weights of the CNN and the optimized value of $\gamma$ are given by those characterizing the model at the end of the early-stopping iteration.

Validation and testing was performed on particles from an independent simulation based on a different realization of the initial density field to those used for training. Although the validation set does not directly enter the training process of the algorithm, it is indirectly used to test the response of the algorithm to changes in the architecture, and to determine the stopping point for training. Validating and testing on independent realizations ensures that the algorithm is not overfitting patterns specific to the simulations used for training. Instead, it ensures that the CNN is learning physical connections between the initial conditions and the final halos which are generalizable to any realization of the initial density field.

\section{A comparison with analytic models}
\label{sec:analytic}
\begin{figure*}
	\includegraphics[width=\textwidth]{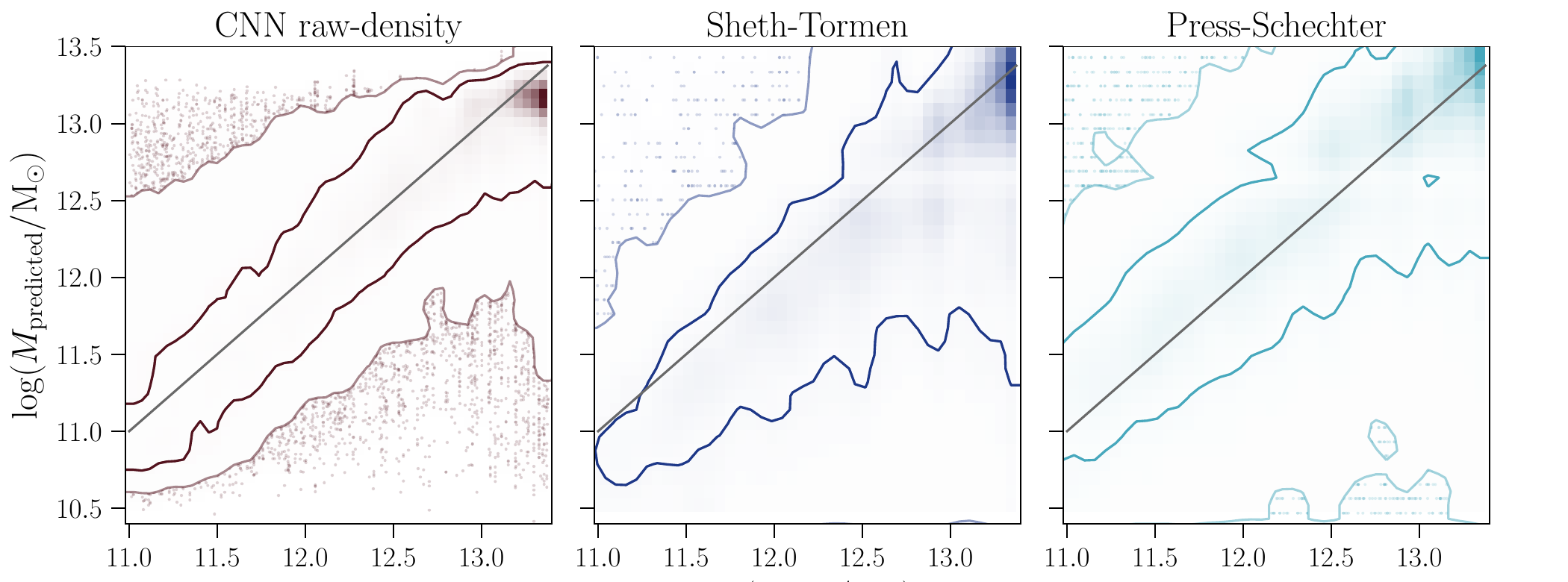}
    \caption{Two-dimensional histograms and contours containing $68\%$ and $95\%$ of the joint probability of the predicted vs. true halo masses for the analytic and CNN models. We compare the raw-density CNN predictions (\textit{left panel}) with those from the Sheth-Tormen (\textit{middle panel}) and extended Press-Schechter (\textit{right panel}) analytic models. The predictions are qualitatively similar, but with tighter confidence regions for the CNN case. This validates our results from the CNN as we find no evidence of any lack of predictive performance from the CNN compared to established analytic models of the same mapping.}
    \label{fig:ML_vs_analytic}
\end{figure*}

We compared the accuracy of the CNN predictions against that of analytic models which also provide final halo mass predictions from the initial conditions. This serves as a validation test for our CNN model. We expect the CNN model to return halo mass predictions that are at least as accurate as those of state-of-the-art analytic approximations, since both CNN models have access to the spherically-averaged density around each particle which is what is used by the analytic models to predict the final halo mass.

We compared the CNN to the extended Press-Schechter (EPS) \cite{Bond1991} and Sheth-Tormen (ST) \cite{ShethTormen2002} analytic halo collapse models. According to EPS, the fraction of density trajectories with a first upcrossing of a density threshold barrier $\delta_\mathrm{th}$ is equivalent to the fraction of haloes of mass $M$. The density threshold barrier $\delta_\mathrm{th}$ adopted by \citet{Bond1991} is that of spherical collapse: $\delta_\mathrm{th}(z) = (D(z)/D(0))\delta_\mathrm{sc}$, where $\delta_\mathrm{sc} \approx 1.686$. The predicted halo mass of each test particle is given by the smoothing mass scale at which the particle first upcrosses the density threshold barrier. 

In the ST formalism, EPS theory is extended by adopting a ``moving" collapse barrier rather than the spherical collapse barrier. The ST collapse barrier $b(z)$ varies as a function of the mass variance $\sigma^2(M)$ and is given by
\begin{equation}
	b(z) = \sqrt{a} \delta_\mathrm{sc}(z) \left[ 1 + \left( \beta \dfrac{\sigma^2 (M)}{a \delta_{\mathrm{sc}}^2 (z)} \right)^{\gamma} \right], 
	\label{eq:ST_barrier}	
\end{equation}
where $\delta_{\mathrm{sc}} (0) \approx 1.686$ and the best-fit parameters found in \citet{ShethMoTormen2001} are $\beta  = 0.485$, $\gamma = 0.615$ and $a = 0.707$. Similar to the EPS case, the predicted halo mass of each test particle is given by the smoothing mass scale at which the particle first upcrosses the threshold barrier given by Eq.~\eqref{eq:ST_barrier}. 
Note that $30\%$ ($58\%$) of particles in the test set have trajectories that never cross the EPS (ST) collapse barrier for the smoothing mass scales we consider, and so these do not have an associated mass prediction. This selection bias will be relevant when making quantitative comparisons with the CNN models in terms of MI, as described below.

\begin{table}
	\setlength{\tabcolsep}{9pt}
	\renewcommand{\arraystretch}{1.5}
	\centering
	\caption{\label{tab:MI_EPS} MI (in nats) between predicted and ground truth halo mass values for the raw-density, averaged-density and EPS models. The MI computation includes only test set particles which have a prediction under the EPS model.}
	\begin{tabular}{ c|c|c}
		\hline
		\textbf{Raw density} & \textbf{Averaged density} & \textbf{EPS}\\ [0.5ex]
        \hline
		0.409 $\pm$ 0.004 & 0.271 $\pm$ 0.004 & 0.257 $\pm$ 0.005 \\
		\hline
	\end{tabular}
\end{table}

\begin{table}
	\setlength{\tabcolsep}{9pt}
	\renewcommand{\arraystretch}{1.5}
	\centering
 	\caption{\label{tab:MI_ST} MI (in nats) between predicted and ground truth halo mass values for the raw-density, averaged-density and ST models. The MI computation includes only test set particles which have a prediction under the ST model.}
    \begin{tabular}{ c|c|c}
		\hline
		\textbf{Raw density} & \textbf{Averaged density} & \textbf{ST }\\ [0.5ex]
        \hline
		0.411 $\pm$ 0.004 & 0.299 $\pm$ 0.004 & 0.260 $\pm$ 0.005\\
		\hline
	\end{tabular}
\end{table}

We computed the EPS and ST predicted halo masses for the particles in the test set used to test the CNN. Fig.~\ref{fig:ML_vs_analytic} shows the predicted halo masses as a function of true halo masses for the analytic and CNN models. We show two-dimensional histograms and the contours containing $68\%$ and $95\%$ of the joint probability. All models show qualitatively similar predictions, but with tighter confidence regions for the CNN predictions. This is especially notable in the bottom-right region of the middle and right panels, where the analytic models' predictions extend to much lower mass values than the CNN predictions. The CNN shows a slight tendency to underpredict the mass of high-mass halos as these are close to the edge of the training set mass range; although our loss function was designed to mitigate this common bias effect in CNNs, it does not entirely remove it. The ST predictions are shifted towards lower mass values compared to the EPS predictions, for fixed true halo mass. This is because the ST collapse barrier takes larger $\delta$ values than the EPS barrier at fixed smoothing mass scale; as a result, the same particle will cross the collapse barrier at lower smoothing mass scales for ST compared to EPS. This in turn yields a lower halo mass prediction for ST compared to EPS.

In Table~\ref{tab:MI_EPS} and \ref{tab:MI_ST}, we quantitatively compare the performance of the CNN models with the analytic ones in terms of the MI between predicted and ground truth halo mass values. When making a quantitative comparison, one must consider that particles whose trajectories do not cross the EPS (or ST) collapse barrier do not  have an associated halo mass prediction. By contrast, the CNN will always return a mass prediction for every particle used for testing. Therefore, we compare the performance of EPS to that of the CNN models for only those subset of particles in the test set that have a predicted mass value according to EPS, and report the MI values in Table~\ref{tab:MI_EPS}. We do the same for ST and report the MI values in Table~\ref{tab:MI_ST}. We find that the raw density and averaged density models perform better than state-of-the-art analytic models for the same set of particles, as the MI of the CNN models is higher than that of the analytic models.

This test validates our results as it confirms that the CNN is at least as accurate as the analytic models.

% Bibliography
\bibliography{dlpaper}

%apsrev4-2.bst 2019-01-14 (MD) hand-edited version of apsrev4-1.bst
%Control: key (0)
%Control: author (8) initials jnrlst
%Control: editor formatted (1) identically to author
%Control: production of article title (0) allowed
%Control: page (0) single
%Control: year (1) truncated
%Control: production of eprint (0) enabled
\begin{thebibliography}{48}%
\makeatletter
\providecommand \@ifxundefined [1]{%
 \@ifx{#1\undefined}
}%
\providecommand \@ifnum [1]{%
 \ifnum #1\expandafter \@firstoftwo
 \else \expandafter \@secondoftwo
 \fi
}%
\providecommand \@ifx [1]{%
 \ifx #1\expandafter \@firstoftwo
 \else \expandafter \@secondoftwo
 \fi
}%
\providecommand \natexlab [1]{#1}%
\providecommand \enquote  [1]{``#1''}%
\providecommand \bibnamefont  [1]{#1}%
\providecommand \bibfnamefont [1]{#1}%
\providecommand \citenamefont [1]{#1}%
\providecommand \href@noop [0]{\@secondoftwo}%
\providecommand \href [0]{\begingroup \@sanitize@url \@href}%
\providecommand \@href[1]{\@@startlink{#1}\@@href}%
\providecommand \@@href[1]{\endgroup#1\@@endlink}%
\providecommand \@sanitize@url [0]{\catcode `\\12\catcode `\$12\catcode
  `\&12\catcode `\#12\catcode `\^12\catcode `\_12\catcode `\%12\relax}%
\providecommand \@@startlink[1]{}%
\providecommand \@@endlink[0]{}%
\providecommand \url  [0]{\begingroup\@sanitize@url \@url }%
\providecommand \@url [1]{\endgroup\@href {#1}{\urlprefix }}%
\providecommand \urlprefix  [0]{URL }%
\providecommand \Eprint [0]{\href }%
\providecommand \doibase [0]{https://doi.org/}%
\providecommand \selectlanguage [0]{\@gobble}%
\providecommand \bibinfo  [0]{\@secondoftwo}%
\providecommand \bibfield  [0]{\@secondoftwo}%
\providecommand \translation [1]{[#1]}%
\providecommand \BibitemOpen [0]{}%
\providecommand \bibitemStop [0]{}%
\providecommand \bibitemNoStop [0]{.\EOS\space}%
\providecommand \EOS [0]{\spacefactor3000\relax}%
\providecommand \BibitemShut  [1]{\csname bibitem#1\endcsname}%
\let\auto@bib@innerbib\@empty
%</preamble>
\bibitem [{\citenamefont {{Efstathiou}}\ \emph {et~al.}(1985)\citenamefont
  {{Efstathiou}}, \citenamefont {{Davis}}, \citenamefont {{White}},\ and\
  \citenamefont {{Frenk}}}]{Efstathiou1985}%
  \BibitemOpen
  \bibfield  {author} {\bibinfo {author} {\bibfnamefont {G.}~\bibnamefont
  {{Efstathiou}}}, \bibinfo {author} {\bibfnamefont {M.}~\bibnamefont
  {{Davis}}}, \bibinfo {author} {\bibfnamefont {S.~D.~M.}\ \bibnamefont
  {{White}}},\ and\ \bibinfo {author} {\bibfnamefont {C.~S.}\ \bibnamefont
  {{Frenk}}},\ }\bibfield  {title} {\bibinfo {title} {{Numerical techniques for
  large cosmological N-body simulations}},\ }\href
  {https://doi.org/10.1086/191003} {\bibfield  {journal} {\bibinfo  {journal}
  {Astrophysical Journal Supplement}\ }\textbf {\bibinfo {volume} {57}},\
  \bibinfo {pages} {241} (\bibinfo {year} {1985})}\BibitemShut {NoStop}%
\bibitem [{\citenamefont {{Jenkins}}\ \emph {et~al.}(2001)\citenamefont
  {{Jenkins}}, \citenamefont {{Frenk}}, \citenamefont {{White}}, \citenamefont
  {{Colberg}}, \citenamefont {{Cole}}, \citenamefont {{Evrard}}, \citenamefont
  {{Couchman}},\ and\ \citenamefont {{Yoshida}}}]{Jenkins2001}%
  \BibitemOpen
  \bibfield  {author} {\bibinfo {author} {\bibfnamefont {A.}~\bibnamefont
  {{Jenkins}}}, \bibinfo {author} {\bibfnamefont {C.~S.}\ \bibnamefont
  {{Frenk}}}, \bibinfo {author} {\bibfnamefont {S.~D.~M.}\ \bibnamefont
  {{White}}}, \bibinfo {author} {\bibfnamefont {J.~M.}\ \bibnamefont
  {{Colberg}}}, \bibinfo {author} {\bibfnamefont {S.}~\bibnamefont {{Cole}}},
  \bibinfo {author} {\bibfnamefont {A.~E.}\ \bibnamefont {{Evrard}}}, \bibinfo
  {author} {\bibfnamefont {H.~M.~P.}\ \bibnamefont {{Couchman}}},\ and\
  \bibinfo {author} {\bibfnamefont {N.}~\bibnamefont {{Yoshida}}},\ }\bibfield
  {title} {\bibinfo {title} {{The mass function of dark matter haloes}},\
  }\href {https://doi.org/10.1046/j.1365-8711.2001.04029.x} {\bibfield
  {journal} {\bibinfo  {journal} {Monthly Notices of the Royal Astronomical
  Society}\ }\textbf {\bibinfo {volume} {321}},\ \bibinfo {pages} {372}
  (\bibinfo {year} {2001})}\BibitemShut {NoStop}%
\bibitem [{\citenamefont {{Navarro}}\ \emph {et~al.}(1997)\citenamefont
  {{Navarro}}, \citenamefont {{Frenk}},\ and\ \citenamefont
  {{White}}}]{Navarro1997}%
  \BibitemOpen
  \bibfield  {author} {\bibinfo {author} {\bibfnamefont {J.~F.}\ \bibnamefont
  {{Navarro}}}, \bibinfo {author} {\bibfnamefont {C.~S.}\ \bibnamefont
  {{Frenk}}},\ and\ \bibinfo {author} {\bibfnamefont {S.~D.~M.}\ \bibnamefont
  {{White}}},\ }\bibfield  {title} {\bibinfo {title} {{A Universal Density
  Profile from Hierarchical Clustering}},\ }\href
  {https://doi.org/10.1086/304888} {\bibfield  {journal} {\bibinfo  {journal}
  {Astrophysical Journal}\ }\textbf {\bibinfo {volume} {490}},\ \bibinfo
  {pages} {493} (\bibinfo {year} {1997})}\BibitemShut {NoStop}%
\bibitem [{\citenamefont {{Springel}}\ \emph {et~al.}(2001)\citenamefont
  {{Springel}}, \citenamefont {{Yoshida}},\ and\ \citenamefont
  {{White}}}]{gadget}%
  \BibitemOpen
  \bibfield  {author} {\bibinfo {author} {\bibfnamefont {V.}~\bibnamefont
  {{Springel}}}, \bibinfo {author} {\bibfnamefont {N.}~\bibnamefont
  {{Yoshida}}},\ and\ \bibinfo {author} {\bibfnamefont {S.~D.~M.}\ \bibnamefont
  {{White}}},\ }\bibfield  {title} {\bibinfo {title} {{GADGET: a code for
  collisionless and gasdynamical cosmological simulations}},\ }\href
  {https://doi.org/10.1016/S1384-1076(01)00042-2} {\bibfield  {journal}
  {\bibinfo  {journal} {New Astronomy}\ }\textbf {\bibinfo {volume} {6}},\
  \bibinfo {pages} {79} (\bibinfo {year} {2001})}\BibitemShut {NoStop}%
\bibitem [{\citenamefont {{Press}}\ and\ \citenamefont
  {{Schechter}}(1974)}]{PressSchechter1974}%
  \BibitemOpen
  \bibfield  {author} {\bibinfo {author} {\bibfnamefont {W.~H.}\ \bibnamefont
  {{Press}}}\ and\ \bibinfo {author} {\bibfnamefont {P.}~\bibnamefont
  {{Schechter}}},\ }\bibfield  {title} {\bibinfo {title} {{Formation of
  Galaxies and Clusters of Galaxies by Self-Similar Gravitational
  Condensation}},\ }\href {https://doi.org/10.1086/152650} {\bibfield
  {journal} {\bibinfo  {journal} {The Astrophysical Journal}\ }\textbf
  {\bibinfo {volume} {187}},\ \bibinfo {pages} {425} (\bibinfo {year}
  {1974})}\BibitemShut {NoStop}%
\bibitem [{\citenamefont {{Bond}}\ \emph {et~al.}(1991)\citenamefont {{Bond}},
  \citenamefont {{Cole}}, \citenamefont {{Efstathiou}},\ and\ \citenamefont
  {{Kaiser}}}]{Bond1991}%
  \BibitemOpen
  \bibfield  {author} {\bibinfo {author} {\bibfnamefont {J.~R.}\ \bibnamefont
  {{Bond}}}, \bibinfo {author} {\bibfnamefont {S.}~\bibnamefont {{Cole}}},
  \bibinfo {author} {\bibfnamefont {G.}~\bibnamefont {{Efstathiou}}},\ and\
  \bibinfo {author} {\bibfnamefont {N.}~\bibnamefont {{Kaiser}}},\ }\bibfield
  {title} {\bibinfo {title} {{Excursion set mass functions for hierarchical
  Gaussian fluctuations}},\ }\href {https://doi.org/10.1086/170520} {\bibfield
  {journal} {\bibinfo  {journal} {The Astrophysical Journal}\ }\textbf
  {\bibinfo {volume} {379}},\ \bibinfo {pages} {440} (\bibinfo {year}
  {1991})}\BibitemShut {NoStop}%
\bibitem [{\citenamefont {{Doroshkevich}}(1970)}]{Doroshkevich1970}%
  \BibitemOpen
  \bibfield  {author} {\bibinfo {author} {\bibfnamefont {A.~G.}\ \bibnamefont
  {{Doroshkevich}}},\ }\bibfield  {title} {\bibinfo {title} {{The space
  structure of perturbations and the origin of rotation of galaxies in the
  theory of fluctuation.}},\ }\href@noop {} {\bibfield  {journal} {\bibinfo
  {journal} {Astrofizika}\ }\textbf {\bibinfo {volume} {6}},\ \bibinfo {pages}
  {581} (\bibinfo {year} {1970})}\BibitemShut {NoStop}%
\bibitem [{\citenamefont {{Bond}}\ and\ \citenamefont
  {{Myers}}(1996{\natexlab{a}})}]{Bond&Myers1996}%
  \BibitemOpen
  \bibfield  {author} {\bibinfo {author} {\bibfnamefont {J.~R.}\ \bibnamefont
  {{Bond}}}\ and\ \bibinfo {author} {\bibfnamefont {S.~T.}\ \bibnamefont
  {{Myers}}},\ }\bibfield  {title} {\bibinfo {title} {{The Peak-Patch Picture
  of Cosmic Catalogs. I. Algorithms}},\ }\href {https://doi.org/10.1086/192267}
  {\bibfield  {journal} {\bibinfo  {journal} {The Astrophysical Journal
  Supplement Series}\ }\textbf {\bibinfo {volume} {103}},\ \bibinfo {pages} {1}
  (\bibinfo {year} {1996}{\natexlab{a}})}\BibitemShut {NoStop}%
\bibitem [{\citenamefont {{Sheth}}\ and\ \citenamefont
  {{Tormen}}(1999)}]{ShethTormen1999}%
  \BibitemOpen
  \bibfield  {author} {\bibinfo {author} {\bibfnamefont {R.~K.}\ \bibnamefont
  {{Sheth}}}\ and\ \bibinfo {author} {\bibfnamefont {G.}~\bibnamefont
  {{Tormen}}},\ }\bibfield  {title} {\bibinfo {title} {{Large-scale bias and
  the peak background split}},\ }\href
  {https://doi.org/10.1046/j.1365-8711.1999.02692.x} {\bibfield  {journal}
  {\bibinfo  {journal} {Monthly Notices of the Royal Astronomical Society}\
  }\textbf {\bibinfo {volume} {308}},\ \bibinfo {pages} {119} (\bibinfo {year}
  {1999})}\BibitemShut {NoStop}%
\bibitem [{\citenamefont {{Sheth}}\ \emph {et~al.}(2001)\citenamefont
  {{Sheth}}, \citenamefont {{Mo}},\ and\ \citenamefont
  {{Tormen}}}]{ShethMoTormen2001}%
  \BibitemOpen
  \bibfield  {author} {\bibinfo {author} {\bibfnamefont {R.~K.}\ \bibnamefont
  {{Sheth}}}, \bibinfo {author} {\bibfnamefont {H.~J.}\ \bibnamefont {{Mo}}},\
  and\ \bibinfo {author} {\bibfnamefont {G.}~\bibnamefont {{Tormen}}},\
  }\bibfield  {title} {\bibinfo {title} {{Ellipsoidal collapse and an improved
  model for the number and spatial distribution of dark matter haloes}},\
  }\href {https://doi.org/10.1046/j.1365-8711.2001.04006.x} {\bibfield
  {journal} {\bibinfo  {journal} {Monthly Notices of the Royal Astronomical
  Society}\ }\textbf {\bibinfo {volume} {323}},\ \bibinfo {pages} {1} (\bibinfo
  {year} {2001})}\BibitemShut {NoStop}%
\bibitem [{\citenamefont {{Sheth}}\ and\ \citenamefont
  {{Tormen}}(2002)}]{ShethTormen2002}%
  \BibitemOpen
  \bibfield  {author} {\bibinfo {author} {\bibfnamefont {R.~K.}\ \bibnamefont
  {{Sheth}}}\ and\ \bibinfo {author} {\bibfnamefont {G.}~\bibnamefont
  {{Tormen}}},\ }\bibfield  {title} {\bibinfo {title} {{An excursion set model
  of hierarchical clustering: ellipsoidal collapse and the moving barrier}},\
  }\href {https://doi.org/10.1046/j.1365-8711.2002.04950.x} {\bibfield
  {journal} {\bibinfo  {journal} {Monthly Notices of the Royal Astronomical
  Society}\ }\textbf {\bibinfo {volume} {329}},\ \bibinfo {pages} {61}
  (\bibinfo {year} {2002})}\BibitemShut {NoStop}%
\bibitem [{\citenamefont {{Lucie-Smith}}\ \emph {et~al.}(2018)\citenamefont
  {{Lucie-Smith}}, \citenamefont {{Peiris}}, \citenamefont {{Pontzen}},\ and\
  \citenamefont {{Lochner}}}]{LucieSmith2018}%
  \BibitemOpen
  \bibfield  {author} {\bibinfo {author} {\bibfnamefont {L.}~\bibnamefont
  {{Lucie-Smith}}}, \bibinfo {author} {\bibfnamefont {H.~V.}\ \bibnamefont
  {{Peiris}}}, \bibinfo {author} {\bibfnamefont {A.}~\bibnamefont
  {{Pontzen}}},\ and\ \bibinfo {author} {\bibfnamefont {M.}~\bibnamefont
  {{Lochner}}},\ }\bibfield  {title} {\bibinfo {title} {{Machine learning
  cosmological structure formation}},\ }\href
  {https://doi.org/10.1093/mnras/sty1719} {\bibfield  {journal} {\bibinfo
  {journal} {Monthly Notices of the Royal Astronomical Society}\ }\textbf
  {\bibinfo {volume} {479}},\ \bibinfo {pages} {3405} (\bibinfo {year}
  {2018})}\BibitemShut {NoStop}%
\bibitem [{\citenamefont {{Lucie-Smith}}\ \emph {et~al.}(2019)\citenamefont
  {{Lucie-Smith}}, \citenamefont {{Peiris}},\ and\ \citenamefont
  {{Pontzen}}}]{LucieSmith2019}%
  \BibitemOpen
  \bibfield  {author} {\bibinfo {author} {\bibfnamefont {L.}~\bibnamefont
  {{Lucie-Smith}}}, \bibinfo {author} {\bibfnamefont {H.~V.}\ \bibnamefont
  {{Peiris}}},\ and\ \bibinfo {author} {\bibfnamefont {A.}~\bibnamefont
  {{Pontzen}}},\ }\bibfield  {title} {\bibinfo {title} {{An interpretable
  machine-learning framework for dark matter halo formation}},\ }\href
  {https://doi.org/10.1093/mnras/stz2599} {\bibfield  {journal} {\bibinfo
  {journal} {Monthly Notices of the Royal Astronomical Society}\ }\textbf
  {\bibinfo {volume} {490}},\ \bibinfo {pages} {331} (\bibinfo {year}
  {2019})}\BibitemShut {NoStop}%
\bibitem [{\citenamefont {LeCun}\ \emph {et~al.}(2015)\citenamefont {LeCun},
  \citenamefont {Bengio},\ and\ \citenamefont {Hinton}}]{LeCun2015}%
  \BibitemOpen
  \bibfield  {author} {\bibinfo {author} {\bibfnamefont {Y.}~\bibnamefont
  {LeCun}}, \bibinfo {author} {\bibfnamefont {Y.}~\bibnamefont {Bengio}},\ and\
  \bibinfo {author} {\bibfnamefont {G.}~\bibnamefont {Hinton}},\ }\bibfield
  {title} {\bibinfo {title} {{Deep learning}},\ }\href
  {https://doi.org/10.1038/nature14539} {\bibfield  {journal} {\bibinfo
  {journal} {Nature}\ }\textbf {\bibinfo {volume} {521}},\ \bibinfo {pages}
  {436} (\bibinfo {year} {2015})}\BibitemShut {NoStop}%
\bibitem [{\citenamefont {Bengio}(2009)}]{Bengio2009deep}%
  \BibitemOpen
  \bibfield  {author} {\bibinfo {author} {\bibfnamefont {Y.}~\bibnamefont
  {Bengio}},\ }\bibfield  {title} {\bibinfo {title} {Learning deep
  architectures for ai},\ }\href {https://doi.org/10.1561/2200000006}
  {\bibfield  {journal} {\bibinfo  {journal} {Foundations and Trends in Machine
  Learning}\ }\textbf {\bibinfo {volume} {2}},\ \bibinfo {pages} {1} (\bibinfo
  {year} {2009})}\BibitemShut {NoStop}%
\bibitem [{\citenamefont {Ravanbakhsh}\ \emph {et~al.}(2016)\citenamefont
  {Ravanbakhsh}, \citenamefont {Oliva}, \citenamefont {Fromenteau},
  \citenamefont {Price}, \citenamefont {Ho}, \citenamefont {Schneider},\ and\
  \citenamefont {P\'{o}czos}}]{Ravanbakhsh2017}%
  \BibitemOpen
  \bibfield  {author} {\bibinfo {author} {\bibfnamefont {S.}~\bibnamefont
  {Ravanbakhsh}}, \bibinfo {author} {\bibfnamefont {J.}~\bibnamefont {Oliva}},
  \bibinfo {author} {\bibfnamefont {S.}~\bibnamefont {Fromenteau}}, \bibinfo
  {author} {\bibfnamefont {L.~C.}\ \bibnamefont {Price}}, \bibinfo {author}
  {\bibfnamefont {S.}~\bibnamefont {Ho}}, \bibinfo {author} {\bibfnamefont
  {J.}~\bibnamefont {Schneider}},\ and\ \bibinfo {author} {\bibfnamefont
  {B.}~\bibnamefont {P\'{o}czos}},\ }\bibfield  {title} {\bibinfo {title}
  {Estimating cosmological parameters from the dark matter distribution},\ }in\
  \href@noop {} {\emph {\bibinfo {booktitle} {Proceedings of the 33rd
  International Conference on International Conference on Machine Learning -
  Volume 48}}},\ \bibinfo {series and number} {ICML'16}\ (\bibinfo  {publisher}
  {JMLR.org},\ \bibinfo {year} {2016})\ p.\ \bibinfo {pages}
  {2407–2416}\BibitemShut {NoStop}%
\bibitem [{\citenamefont {Mathuriya}\ \emph {et~al.}(2018)\citenamefont
  {Mathuriya}, \citenamefont {Bard}, \citenamefont {Mendygral}, \citenamefont
  {Meadows}, \citenamefont {Arnemann}, \citenamefont {Shao}, \citenamefont
  {He}, \citenamefont {K\"{a}rn\"{a}}, \citenamefont {Moise}, \citenamefont
  {Pennycook}, \citenamefont {Maschhoff}, \citenamefont {Sewall}, \citenamefont
  {Kumar}, \citenamefont {Ho}, \citenamefont {Ringenburg}, \citenamefont
  {Prabhat},\ and\ \citenamefont {Lee}}]{Mathuriya2018}%
  \BibitemOpen
  \bibfield  {author} {\bibinfo {author} {\bibfnamefont {A.}~\bibnamefont
  {Mathuriya}}, \bibinfo {author} {\bibfnamefont {D.}~\bibnamefont {Bard}},
  \bibinfo {author} {\bibfnamefont {P.}~\bibnamefont {Mendygral}}, \bibinfo
  {author} {\bibfnamefont {L.}~\bibnamefont {Meadows}}, \bibinfo {author}
  {\bibfnamefont {J.}~\bibnamefont {Arnemann}}, \bibinfo {author}
  {\bibfnamefont {L.}~\bibnamefont {Shao}}, \bibinfo {author} {\bibfnamefont
  {S.}~\bibnamefont {He}}, \bibinfo {author} {\bibfnamefont {T.}~\bibnamefont
  {K\"{a}rn\"{a}}}, \bibinfo {author} {\bibfnamefont {D.}~\bibnamefont
  {Moise}}, \bibinfo {author} {\bibfnamefont {S.~J.}\ \bibnamefont
  {Pennycook}}, \bibinfo {author} {\bibfnamefont {K.}~\bibnamefont
  {Maschhoff}}, \bibinfo {author} {\bibfnamefont {J.}~\bibnamefont {Sewall}},
  \bibinfo {author} {\bibfnamefont {N.}~\bibnamefont {Kumar}}, \bibinfo
  {author} {\bibfnamefont {S.}~\bibnamefont {Ho}}, \bibinfo {author}
  {\bibfnamefont {M.~F.}\ \bibnamefont {Ringenburg}}, \bibinfo {author}
  {\bibnamefont {Prabhat}},\ and\ \bibinfo {author} {\bibfnamefont
  {V.}~\bibnamefont {Lee}},\ }\bibfield  {title} {\bibinfo {title} {Cosmoflow:
  Using deep learning to learn the universe at scale},\ }in\ \href
  {https://doi.org/10.1109/SC.2018.00068} {\emph {\bibinfo {booktitle}
  {Proceedings of the International Conference for High Performance Computing,
  Networking, Storage, and Analysis}}},\ \bibinfo {series and number} {SC '18}\
  (\bibinfo  {publisher} {IEEE Press},\ \bibinfo {year} {2018})\BibitemShut
  {NoStop}%
\bibitem [{\citenamefont {{Pan}}\ \emph {et~al.}(2020)\citenamefont {{Pan}},
  \citenamefont {{Liu}}, \citenamefont {{Forero-Romero}}, \citenamefont
  {{Sabiu}}, \citenamefont {{Li}}, \citenamefont {{Miao}},\ and\ \citenamefont
  {{Li}}}]{Pan2020}%
  \BibitemOpen
  \bibfield  {author} {\bibinfo {author} {\bibfnamefont {S.}~\bibnamefont
  {{Pan}}}, \bibinfo {author} {\bibfnamefont {M.}~\bibnamefont {{Liu}}},
  \bibinfo {author} {\bibfnamefont {J.}~\bibnamefont {{Forero-Romero}}},
  \bibinfo {author} {\bibfnamefont {C.~G.}\ \bibnamefont {{Sabiu}}}, \bibinfo
  {author} {\bibfnamefont {Z.}~\bibnamefont {{Li}}}, \bibinfo {author}
  {\bibfnamefont {H.}~\bibnamefont {{Miao}}},\ and\ \bibinfo {author}
  {\bibfnamefont {X.-D.}\ \bibnamefont {{Li}}},\ }\bibfield  {title} {\bibinfo
  {title} {{Cosmological parameter estimation from large-scale structure deep
  learning}},\ }\href {https://doi.org/10.1007/s11433-020-1586-3} {\bibfield
  {journal} {\bibinfo  {journal} {Science China Physics, Mechanics, and
  Astronomy}\ }\textbf {\bibinfo {volume} {63}},\ \bibinfo {eid} {110412}
  (\bibinfo {year} {2020})}\BibitemShut {NoStop}%
\bibitem [{\citenamefont {{Villaescusa-Navarro}}\ \emph
  {et~al.}(2020)\citenamefont {{Villaescusa-Navarro}}, \citenamefont
  {{Wandelt}}, \citenamefont {{Angl{\'e}s-Alc{\'a}zar}}, \citenamefont
  {{Genel}}, \citenamefont {{Zorrilla Mantilla}}, \citenamefont {{Ho}},\ and\
  \citenamefont {{Spergel}}}]{Villaescusa-Navarro2020}%
  \BibitemOpen
  \bibfield  {author} {\bibinfo {author} {\bibfnamefont {F.}~\bibnamefont
  {{Villaescusa-Navarro}}}, \bibinfo {author} {\bibfnamefont {B.~D.}\
  \bibnamefont {{Wandelt}}}, \bibinfo {author} {\bibfnamefont {D.}~\bibnamefont
  {{Angl{\'e}s-Alc{\'a}zar}}}, \bibinfo {author} {\bibfnamefont
  {S.}~\bibnamefont {{Genel}}}, \bibinfo {author} {\bibfnamefont {J.~M.}\
  \bibnamefont {{Zorrilla Mantilla}}}, \bibinfo {author} {\bibfnamefont
  {S.}~\bibnamefont {{Ho}}},\ and\ \bibinfo {author} {\bibfnamefont {D.~N.}\
  \bibnamefont {{Spergel}}},\ }\bibfield  {title} {\bibinfo {title} {{Neural
  networks as optimal estimators to marginalize over baryonic effects.}},\
  }\href@noop {} {\bibfield  {journal} {\bibinfo  {journal} {arXiv}\ ,\
  \bibinfo {pages} {Preprint at arXiv:2011.05992}} (\bibinfo {year}
  {2020})}\BibitemShut {NoStop}%
\bibitem [{\citenamefont {{Ntampaka}}\ \emph {et~al.}(2020)\citenamefont
  {{Ntampaka}}, \citenamefont {{Eisenstein}}, \citenamefont {{Yuan}},\ and\
  \citenamefont {{Garrison}}}]{Ntampaka2020}%
  \BibitemOpen
  \bibfield  {author} {\bibinfo {author} {\bibfnamefont {M.}~\bibnamefont
  {{Ntampaka}}}, \bibinfo {author} {\bibfnamefont {D.~J.}\ \bibnamefont
  {{Eisenstein}}}, \bibinfo {author} {\bibfnamefont {S.}~\bibnamefont
  {{Yuan}}},\ and\ \bibinfo {author} {\bibfnamefont {L.~H.}\ \bibnamefont
  {{Garrison}}},\ }\bibfield  {title} {\bibinfo {title} {{A Hybrid Deep
  Learning Approach to Cosmological Constraints from Galaxy Redshift
  Surveys}},\ }\href {https://doi.org/10.3847/1538-4357/ab5f5e} {\bibfield
  {journal} {\bibinfo  {journal} {Astrophysical Journal}\ }\textbf {\bibinfo
  {volume} {889}},\ \bibinfo {eid} {151} (\bibinfo {year} {2020})}\BibitemShut
  {NoStop}%
\bibitem [{\citenamefont {Moster}\ \emph {et~al.}(2021)\citenamefont {Moster},
  \citenamefont {Naab}, \citenamefont {Lindström},\ and\ \citenamefont
  {O’Leary}}]{Moster2020}%
  \BibitemOpen
  \bibfield  {author} {\bibinfo {author} {\bibfnamefont {B.~P.}\ \bibnamefont
  {Moster}}, \bibinfo {author} {\bibfnamefont {T.}~\bibnamefont {Naab}},
  \bibinfo {author} {\bibfnamefont {M.}~\bibnamefont {Lindström}},\ and\
  \bibinfo {author} {\bibfnamefont {J.~A.}\ \bibnamefont {O’Leary}},\
  }\bibfield  {title} {\bibinfo {title} {Galaxynet: connecting galaxies and
  dark matter haloes with deep neural networks and reinforcement learning in
  large volumes},\ }\href {https://doi.org/10.1093/mnras/stab1449} {\bibfield
  {journal} {\bibinfo  {journal} {Monthly Notices of the Royal Astronomical
  Society}\ }\textbf {\bibinfo {volume} {507}},\ \bibinfo {pages} {2115–2136}
  (\bibinfo {year} {2021})}\BibitemShut {NoStop}%
\bibitem [{\citenamefont {Kodi~Ramanah}\ \emph {et~al.}(2020)\citenamefont
  {Kodi~Ramanah}, \citenamefont {Charnock}, \citenamefont
  {Villaescusa-Navarro},\ and\ \citenamefont {Wandelt}}]{KodiRamanah2020}%
  \BibitemOpen
  \bibfield  {author} {\bibinfo {author} {\bibfnamefont {D.}~\bibnamefont
  {Kodi~Ramanah}}, \bibinfo {author} {\bibfnamefont {T.}~\bibnamefont
  {Charnock}}, \bibinfo {author} {\bibfnamefont {F.}~\bibnamefont
  {Villaescusa-Navarro}},\ and\ \bibinfo {author} {\bibfnamefont {B.~D.}\
  \bibnamefont {Wandelt}},\ }\bibfield  {title} {\bibinfo {title}
  {Super-resolution emulator of cosmological simulations using deep physical
  models},\ }\href {https://doi.org/10.1093/mnras/staa1428} {\bibfield
  {journal} {\bibinfo  {journal} {Monthly Notices of the Royal Astronomical
  Society}\ }\textbf {\bibinfo {volume} {495}},\ \bibinfo {pages} {4227–4236}
  (\bibinfo {year} {2020})}\BibitemShut {NoStop}%
\bibitem [{\citenamefont {Li}\ \emph {et~al.}(2021)\citenamefont {Li},
  \citenamefont {Ni}, \citenamefont {Croft}, \citenamefont {Di~Matteo},
  \citenamefont {Bird},\ and\ \citenamefont {Feng}}]{Li2021}%
  \BibitemOpen
  \bibfield  {author} {\bibinfo {author} {\bibfnamefont {Y.}~\bibnamefont
  {Li}}, \bibinfo {author} {\bibfnamefont {Y.}~\bibnamefont {Ni}}, \bibinfo
  {author} {\bibfnamefont {R.~A.~C.}\ \bibnamefont {Croft}}, \bibinfo {author}
  {\bibfnamefont {T.}~\bibnamefont {Di~Matteo}}, \bibinfo {author}
  {\bibfnamefont {S.}~\bibnamefont {Bird}},\ and\ \bibinfo {author}
  {\bibfnamefont {Y.}~\bibnamefont {Feng}},\ }\bibfield  {title} {\bibinfo
  {title} {Ai-assisted superresolution cosmological simulations},\ }\href
  {https://doi.org/10.1073/pnas.2022038118} {\bibfield  {journal} {\bibinfo
  {journal} {Proceedings of the National Academy of Sciences}\ }\textbf
  {\bibinfo {volume} {118}},\ \bibinfo {pages} {e2022038118} (\bibinfo {year}
  {2021})}\BibitemShut {NoStop}%
\bibitem [{\citenamefont {{He}}\ \emph {et~al.}(2019)\citenamefont {{He}},
  \citenamefont {{Li}}, \citenamefont {{Feng}}, \citenamefont {{Ho}},
  \citenamefont {{Ravanbakhsh}}, \citenamefont {{Chen}},\ and\ \citenamefont
  {{P{\'o}czos}}}]{He2019}%
  \BibitemOpen
  \bibfield  {author} {\bibinfo {author} {\bibfnamefont {S.}~\bibnamefont
  {{He}}}, \bibinfo {author} {\bibfnamefont {Y.}~\bibnamefont {{Li}}}, \bibinfo
  {author} {\bibfnamefont {Y.}~\bibnamefont {{Feng}}}, \bibinfo {author}
  {\bibfnamefont {S.}~\bibnamefont {{Ho}}}, \bibinfo {author} {\bibfnamefont
  {S.}~\bibnamefont {{Ravanbakhsh}}}, \bibinfo {author} {\bibfnamefont
  {W.}~\bibnamefont {{Chen}}},\ and\ \bibinfo {author} {\bibfnamefont
  {B.}~\bibnamefont {{P{\'o}czos}}},\ }\bibfield  {title} {\bibinfo {title}
  {{Learning to predict the cosmological structure formation}},\ }\href
  {https://doi.org/10.1073/pnas.1821458116} {\bibfield  {journal} {\bibinfo
  {journal} {Proceedings of the National Academy of Science}\ }\textbf
  {\bibinfo {volume} {116}},\ \bibinfo {pages} {13825} (\bibinfo {year}
  {2019})}\BibitemShut {NoStop}%
\bibitem [{\citenamefont {{Zhang}}\ \emph {et~al.}(2019)\citenamefont
  {{Zhang}}, \citenamefont {{Wang}}, \citenamefont {{Zhang}}, \citenamefont
  {{Sun}}, \citenamefont {{He}}, \citenamefont {{Contardo}}, \citenamefont
  {{Villaescusa-Navarro}},\ and\ \citenamefont {{Ho}}}]{Zhang2019}%
  \BibitemOpen
  \bibfield  {author} {\bibinfo {author} {\bibfnamefont {X.}~\bibnamefont
  {{Zhang}}}, \bibinfo {author} {\bibfnamefont {Y.}~\bibnamefont {{Wang}}},
  \bibinfo {author} {\bibfnamefont {W.}~\bibnamefont {{Zhang}}}, \bibinfo
  {author} {\bibfnamefont {Y.}~\bibnamefont {{Sun}}}, \bibinfo {author}
  {\bibfnamefont {S.}~\bibnamefont {{He}}}, \bibinfo {author} {\bibfnamefont
  {G.}~\bibnamefont {{Contardo}}}, \bibinfo {author} {\bibfnamefont
  {F.}~\bibnamefont {{Villaescusa-Navarro}}},\ and\ \bibinfo {author}
  {\bibfnamefont {S.}~\bibnamefont {{Ho}}},\ }\bibfield  {title} {\bibinfo
  {title} {{From Dark Matter to Galaxies with Convolutional Networks.}},\
  }\href@noop {} {\bibfield  {journal} {\bibinfo  {journal} {arXiv}\ ,\
  \bibinfo {eid} {arXiv:1902.05965}} (\bibinfo {year} {2019})}\BibitemShut
  {NoStop}%
\bibitem [{\citenamefont {{Kodi Ramanah}}\ \emph {et~al.}(2019)\citenamefont
  {{Kodi Ramanah}}, \citenamefont {{Charnock}},\ and\ \citenamefont
  {{Lavaux}}}]{Ramanah2019}%
  \BibitemOpen
  \bibfield  {author} {\bibinfo {author} {\bibfnamefont {D.}~\bibnamefont
  {{Kodi Ramanah}}}, \bibinfo {author} {\bibfnamefont {T.}~\bibnamefont
  {{Charnock}}},\ and\ \bibinfo {author} {\bibfnamefont {G.}~\bibnamefont
  {{Lavaux}}},\ }\bibfield  {title} {\bibinfo {title} {{Painting halos from
  cosmic density fields of dark matter with physically motivated neural
  networks}},\ }\href {https://doi.org/10.1103/PhysRevD.100.043515} {\bibfield
  {journal} {\bibinfo  {journal} {Physical Review D}\ }\textbf {\bibinfo
  {volume} {100}},\ \bibinfo {eid} {043515} (\bibinfo {year}
  {2019})}\BibitemShut {NoStop}%
\bibitem [{\citenamefont {{Springel}}(2005)}]{gadget2}%
  \BibitemOpen
  \bibfield  {author} {\bibinfo {author} {\bibfnamefont {V.}~\bibnamefont
  {{Springel}}},\ }\bibfield  {title} {\bibinfo {title} {{The cosmological
  simulation code GADGET-2}},\ }\href
  {https://doi.org/10.1111/j.1365-2966.2005.09655.x} {\bibfield  {journal}
  {\bibinfo  {journal} {Monthly Notices of the Royal Astronomical Society}\
  }\textbf {\bibinfo {volume} {364}},\ \bibinfo {pages} {1105} (\bibinfo {year}
  {2005})}\BibitemShut {NoStop}%
\bibitem [{\citenamefont {{Pontzen}}\ \emph {et~al.}(2013)\citenamefont
  {{Pontzen}}, \citenamefont {{Ro{\v s}kar}}, \citenamefont {{Stinson}},\ and\
  \citenamefont {{Woods}}}]{pynbody}%
  \BibitemOpen
  \bibfield  {author} {\bibinfo {author} {\bibfnamefont {A.}~\bibnamefont
  {{Pontzen}}}, \bibinfo {author} {\bibfnamefont {R.}~\bibnamefont {{Ro{\v
  s}kar}}}, \bibinfo {author} {\bibfnamefont {G.}~\bibnamefont {{Stinson}}},\
  and\ \bibinfo {author} {\bibfnamefont {R.}~\bibnamefont {{Woods}}},\
  }\href@noop {} {\bibinfo {title} {{pynbody: N-Body/SPH analysis for
  python}}},\ \bibinfo {howpublished} {Astrophysics Source Code Library}
  (\bibinfo {year} {2013})\BibitemShut {NoStop}%
\bibitem [{\citenamefont {{Dunkley}}\ \emph {et~al.}(2009)\citenamefont
  {{Dunkley}}, \citenamefont {{Komatsu}}, \citenamefont {{Nolta}},
  \citenamefont {{Spergel}}, \citenamefont {{Larson}}, \citenamefont
  {{Hinshaw}}, \citenamefont {{Page}}, \citenamefont {{Bennett}}, \citenamefont
  {{Gold}}, \citenamefont {{Jarosik}}, \citenamefont {{Weiland}}, \citenamefont
  {{Halpern}}, \citenamefont {{Hill}}, \citenamefont {{Kogut}}, \citenamefont
  {{Limon}}, \citenamefont {{Meyer}}, \citenamefont {{Tucker}}, \citenamefont
  {{Wollack}},\ and\ \citenamefont {{Wright}}}]{WMAP}%
  \BibitemOpen
  \bibfield  {author} {\bibinfo {author} {\bibfnamefont {J.}~\bibnamefont
  {{Dunkley}}}, \bibinfo {author} {\bibfnamefont {E.}~\bibnamefont
  {{Komatsu}}}, \bibinfo {author} {\bibfnamefont {M.~R.}\ \bibnamefont
  {{Nolta}}}, \bibinfo {author} {\bibfnamefont {D.~N.}\ \bibnamefont
  {{Spergel}}}, \bibinfo {author} {\bibfnamefont {D.}~\bibnamefont {{Larson}}},
  \bibinfo {author} {\bibfnamefont {G.}~\bibnamefont {{Hinshaw}}}, \bibinfo
  {author} {\bibfnamefont {L.}~\bibnamefont {{Page}}}, \bibinfo {author}
  {\bibfnamefont {C.~L.}\ \bibnamefont {{Bennett}}}, \bibinfo {author}
  {\bibfnamefont {B.}~\bibnamefont {{Gold}}}, \bibinfo {author} {\bibfnamefont
  {N.}~\bibnamefont {{Jarosik}}}, \bibinfo {author} {\bibfnamefont {J.~L.}\
  \bibnamefont {{Weiland}}}, \bibinfo {author} {\bibfnamefont {M.}~\bibnamefont
  {{Halpern}}}, \bibinfo {author} {\bibfnamefont {R.~S.}\ \bibnamefont
  {{Hill}}}, \bibinfo {author} {\bibfnamefont {A.}~\bibnamefont {{Kogut}}},
  \bibinfo {author} {\bibfnamefont {M.}~\bibnamefont {{Limon}}}, \bibinfo
  {author} {\bibfnamefont {S.~S.}\ \bibnamefont {{Meyer}}}, \bibinfo {author}
  {\bibfnamefont {G.~S.}\ \bibnamefont {{Tucker}}}, \bibinfo {author}
  {\bibfnamefont {E.}~\bibnamefont {{Wollack}}},\ and\ \bibinfo {author}
  {\bibfnamefont {E.~L.}\ \bibnamefont {{Wright}}},\ }\bibfield  {title}
  {\bibinfo {title} {{Five-Year Wilkinson Microwave Anisotropy Probe
  Observations: Likelihoods and Parameters from the WMAP Data}},\ }\href
  {https://doi.org/10.1088/0067-0049/180/2/306} {\bibfield  {journal} {\bibinfo
   {journal} {The Astrophysical Journal Supplement Series}\ }\textbf {\bibinfo
  {volume} {180}},\ \bibinfo {pages} {306} (\bibinfo {year}
  {2009})}\BibitemShut {NoStop}%
\bibitem [{\citenamefont {{Planck Collaboration}}\ and\ \citenamefont
  {{others}}(2020)}]{Aghanim2020}%
  \BibitemOpen
  \bibfield  {author} {\bibinfo {author} {\bibnamefont {{Planck
  Collaboration}}}\ and\ \bibinfo {author} {\bibnamefont {{others}}},\
  }\bibfield  {title} {\bibinfo {title} {{Planck 2018 results. VI. Cosmological
  parameters}},\ }\href {https://doi.org/10.1051/0004-6361/201833910}
  {\bibfield  {journal} {\bibinfo  {journal} {Astronomy and Astrophysics}\
  }\textbf {\bibinfo {volume} {641}},\ \bibinfo {eid} {A6} (\bibinfo {year}
  {2020})}\BibitemShut {NoStop}%
\bibitem [{\citenamefont {Stopyra}\ \emph {et~al.}(2021)\citenamefont
  {Stopyra}, \citenamefont {Pontzen}, \citenamefont {Peiris}, \citenamefont
  {Roth},\ and\ \citenamefont {Rey}}]{Stopyra2020}%
  \BibitemOpen
  \bibfield  {author} {\bibinfo {author} {\bibfnamefont {S.}~\bibnamefont
  {Stopyra}}, \bibinfo {author} {\bibfnamefont {A.}~\bibnamefont {Pontzen}},
  \bibinfo {author} {\bibfnamefont {H.}~\bibnamefont {Peiris}}, \bibinfo
  {author} {\bibfnamefont {N.}~\bibnamefont {Roth}},\ and\ \bibinfo {author}
  {\bibfnamefont {M.~P.}\ \bibnamefont {Rey}},\ }\bibfield  {title} {\bibinfo
  {title} {Genetic—a new initial conditions generator to support genetically
  modified zoom simulations},\ }\href
  {https://doi.org/10.3847/1538-4365/abcd94} {\bibfield  {journal} {\bibinfo
  {journal} {The Astrophysical Journal Supplement Series}\ }\textbf {\bibinfo
  {volume} {252}},\ \bibinfo {pages} {28} (\bibinfo {year} {2021})}\BibitemShut
  {NoStop}%
\bibitem [{\citenamefont {Nair}\ and\ \citenamefont {Hinton}(2010)}]{Nair2010}%
  \BibitemOpen
  \bibfield  {author} {\bibinfo {author} {\bibfnamefont {V.}~\bibnamefont
  {Nair}}\ and\ \bibinfo {author} {\bibfnamefont {G.~E.}\ \bibnamefont
  {Hinton}},\ }\bibfield  {title} {\bibinfo {title} {Rectified linear units
  improve restricted boltzmann machines},\ }in\ \href
  {http://dl.acm.org/citation.cfm?id=3104322.3104425} {\emph {\bibinfo
  {booktitle} {Proceedings of the 27th International Conference on
  International Conference on Machine Learning}}},\ \bibinfo {series and
  number} {ICML'10}\ (\bibinfo  {publisher} {Omnipress},\ \bibinfo {address}
  {USA},\ \bibinfo {year} {2010})\ pp.\ \bibinfo {pages} {807--814}\BibitemShut
  {NoStop}%
\bibitem [{\citenamefont {{Rumelhart}}\ \emph {et~al.}(1986)\citenamefont
  {{Rumelhart}}, \citenamefont {{Hinton}},\ and\ \citenamefont
  {{Williams}}}]{Rumelhart1986}%
  \BibitemOpen
  \bibfield  {author} {\bibinfo {author} {\bibfnamefont {D.~E.}\ \bibnamefont
  {{Rumelhart}}}, \bibinfo {author} {\bibfnamefont {G.~E.}\ \bibnamefont
  {{Hinton}}},\ and\ \bibinfo {author} {\bibfnamefont {R.~J.}\ \bibnamefont
  {{Williams}}},\ }\bibfield  {title} {\bibinfo {title} {{Learning
  representations by back-propagating errors}},\ }\href
  {https://doi.org/10.1038/323533a0} {\bibfield  {journal} {\bibinfo  {journal}
  {Nature}\ }\textbf {\bibinfo {volume} {323}},\ \bibinfo {pages} {533}
  (\bibinfo {year} {1986})}\BibitemShut {NoStop}%
\bibitem [{\citenamefont {J.~Reddi}\ \emph {et~al.}(2018)\citenamefont
  {J.~Reddi}, \citenamefont {Kale},\ and\ \citenamefont {Kumar}}]{Reddi2019}%
  \BibitemOpen
  \bibfield  {author} {\bibinfo {author} {\bibfnamefont {S.}~\bibnamefont
  {J.~Reddi}}, \bibinfo {author} {\bibfnamefont {S.}~\bibnamefont {Kale}},\
  and\ \bibinfo {author} {\bibfnamefont {S.}~\bibnamefont {Kumar}},\ }\bibfield
   {title} {\bibinfo {title} {On the convergence of adam \& beyond}\ }(\bibinfo
  {year} {2018})\BibitemShut {NoStop}%
\bibitem [{\citenamefont {Kingma}\ and\ \citenamefont {Ba}(2014)}]{Kingma2014}%
  \BibitemOpen
  \bibfield  {author} {\bibinfo {author} {\bibfnamefont {D.}~\bibnamefont
  {Kingma}}\ and\ \bibinfo {author} {\bibfnamefont {J.}~\bibnamefont {Ba}},\
  }\bibfield  {title} {\bibinfo {title} {Adam: A method for stochastic
  optimization},\ }\href@noop {} {\bibfield  {journal} {\bibinfo  {journal}
  {International Conference on Learning Representations}\ } (\bibinfo {year}
  {2014})}\BibitemShut {NoStop}%
\bibitem [{\citenamefont {Piras}\ \emph {et~al.}(2023)\citenamefont {Piras},
  \citenamefont {Peiris}, \citenamefont {Pontzen}, \citenamefont {Lucie-Smith},
  \citenamefont {Guo},\ and\ \citenamefont {Nord}}]{Piras2022}%
  \BibitemOpen
  \bibfield  {author} {\bibinfo {author} {\bibfnamefont {D.}~\bibnamefont
  {Piras}}, \bibinfo {author} {\bibfnamefont {H.~V.}\ \bibnamefont {Peiris}},
  \bibinfo {author} {\bibfnamefont {A.}~\bibnamefont {Pontzen}}, \bibinfo
  {author} {\bibfnamefont {L.}~\bibnamefont {Lucie-Smith}}, \bibinfo {author}
  {\bibfnamefont {N.}~\bibnamefont {Guo}},\ and\ \bibinfo {author}
  {\bibfnamefont {B.}~\bibnamefont {Nord}},\ }\bibfield  {title} {\bibinfo
  {title} {A robust estimator of mutual information for deep learning
  interpretability},\ }\href {https://doi.org/10.1088/2632-2153/acc444}
  {\bibfield  {journal} {\bibinfo  {journal} {Machine Learning: Science and
  Technology}\ }\textbf {\bibinfo {volume} {4}},\ \bibinfo {pages} {025006}
  (\bibinfo {year} {2023})}\BibitemShut {NoStop}%
\bibitem [{\citenamefont {{Szegedy}}\ \emph {et~al.}(2013)\citenamefont
  {{Szegedy}}, \citenamefont {{Zaremba}}, \citenamefont {{Sutskever}},
  \citenamefont {{Bruna}}, \citenamefont {{Erhan}}, \citenamefont
  {{Goodfellow}},\ and\ \citenamefont {{Fergus}}}]{Szegedy2013}%
  \BibitemOpen
  \bibfield  {author} {\bibinfo {author} {\bibfnamefont {C.}~\bibnamefont
  {{Szegedy}}}, \bibinfo {author} {\bibfnamefont {W.}~\bibnamefont
  {{Zaremba}}}, \bibinfo {author} {\bibfnamefont {I.}~\bibnamefont
  {{Sutskever}}}, \bibinfo {author} {\bibfnamefont {J.}~\bibnamefont
  {{Bruna}}}, \bibinfo {author} {\bibfnamefont {D.}~\bibnamefont {{Erhan}}},
  \bibinfo {author} {\bibfnamefont {I.}~\bibnamefont {{Goodfellow}}},\ and\
  \bibinfo {author} {\bibfnamefont {R.}~\bibnamefont {{Fergus}}},\ }\bibfield
  {title} {\bibinfo {title} {{Intriguing properties of neural networks}},\
  }\href@noop {} {\bibfield  {journal} {\bibinfo  {journal} {arXiv e-prints}\
  ,\ \bibinfo {eid} {arXiv:1312.6199}} (\bibinfo {year} {2013})},\ \Eprint
  {https://arxiv.org/abs/1312.6199} {arXiv:1312.6199 [cs.CV]} \BibitemShut
  {NoStop}%
\bibitem [{\citenamefont {{Fluri}}\ \emph {et~al.}(2018)\citenamefont
  {{Fluri}}, \citenamefont {{Kacprzak}}, \citenamefont {{Refregier}},
  \citenamefont {{Amara}}, \citenamefont {{Lucchi}},\ and\ \citenamefont
  {{Hofmann}}}]{Fluri2018}%
  \BibitemOpen
  \bibfield  {author} {\bibinfo {author} {\bibfnamefont {J.}~\bibnamefont
  {{Fluri}}}, \bibinfo {author} {\bibfnamefont {T.}~\bibnamefont {{Kacprzak}}},
  \bibinfo {author} {\bibfnamefont {A.}~\bibnamefont {{Refregier}}}, \bibinfo
  {author} {\bibfnamefont {A.}~\bibnamefont {{Amara}}}, \bibinfo {author}
  {\bibfnamefont {A.}~\bibnamefont {{Lucchi}}},\ and\ \bibinfo {author}
  {\bibfnamefont {T.}~\bibnamefont {{Hofmann}}},\ }\bibfield  {title} {\bibinfo
  {title} {{Cosmological constraints from noisy convergence maps through deep
  learning}},\ }\href {https://doi.org/10.1103/PhysRevD.98.123518} {\bibfield
  {journal} {\bibinfo  {journal} {\prd}\ }\textbf {\bibinfo {volume} {98}},\
  \bibinfo {eid} {123518} (\bibinfo {year} {2018})},\ \Eprint
  {https://arxiv.org/abs/1807.08732} {arXiv:1807.08732 [astro-ph.CO]}
  \BibitemShut {NoStop}%
\bibitem [{\citenamefont {{Schmelzle}}\ \emph {et~al.}(2017)\citenamefont
  {{Schmelzle}}, \citenamefont {{Lucchi}}, \citenamefont {{Kacprzak}},
  \citenamefont {{Amara}}, \citenamefont {{Sgier}}, \citenamefont
  {{R{\'e}fr{\'e}gier}},\ and\ \citenamefont {{Hofmann}}}]{Schmelzle2017}%
  \BibitemOpen
  \bibfield  {author} {\bibinfo {author} {\bibfnamefont {J.}~\bibnamefont
  {{Schmelzle}}}, \bibinfo {author} {\bibfnamefont {A.}~\bibnamefont
  {{Lucchi}}}, \bibinfo {author} {\bibfnamefont {T.}~\bibnamefont
  {{Kacprzak}}}, \bibinfo {author} {\bibfnamefont {A.}~\bibnamefont {{Amara}}},
  \bibinfo {author} {\bibfnamefont {R.}~\bibnamefont {{Sgier}}}, \bibinfo
  {author} {\bibfnamefont {A.}~\bibnamefont {{R{\'e}fr{\'e}gier}}},\ and\
  \bibinfo {author} {\bibfnamefont {T.}~\bibnamefont {{Hofmann}}},\ }\bibfield
  {title} {\bibinfo {title} {{Cosmological model discrimination with Deep
  Learning}},\ }\href@noop {} {\bibfield  {journal} {\bibinfo  {journal} {arXiv
  e-prints}\ ,\ \bibinfo {eid} {arXiv:1707.05167}} (\bibinfo {year} {2017})},\
  \Eprint {https://arxiv.org/abs/1707.05167} {arXiv:1707.05167 [astro-ph.CO]}
  \BibitemShut {NoStop}%
\bibitem [{\citenamefont {Monaco}\ \emph {et~al.}(2002)\citenamefont {Monaco},
  \citenamefont {Theuns},\ and\ \citenamefont {Taffoni}}]{Monaco2002}%
  \BibitemOpen
  \bibfield  {author} {\bibinfo {author} {\bibfnamefont {P.}~\bibnamefont
  {Monaco}}, \bibinfo {author} {\bibfnamefont {T.}~\bibnamefont {Theuns}},\
  and\ \bibinfo {author} {\bibfnamefont {G.}~\bibnamefont {Taffoni}},\
  }\bibfield  {title} {\bibinfo {title} {The pinocchio algorithm: pinpointing
  orbit-crossing collapsed hierarchical objects in a linear density field},\
  }\href {https://doi.org/10.1046/j.1365-8711.2002.05162.x} {\bibfield
  {journal} {\bibinfo  {journal} {Monthly Notices of the Royal Astronomical
  Society}\ }\textbf {\bibinfo {volume} {331}},\ \bibinfo {pages} {587–608}
  (\bibinfo {year} {2002})}\BibitemShut {NoStop}%
\bibitem [{\citenamefont {{Castorina}}\ \emph {et~al.}(2016)\citenamefont
  {{Castorina}}, \citenamefont {{Paranjape}}, \citenamefont {{Hahn}},\ and\
  \citenamefont {{Sheth}}}]{castorina2016}%
  \BibitemOpen
  \bibfield  {author} {\bibinfo {author} {\bibfnamefont {E.}~\bibnamefont
  {{Castorina}}}, \bibinfo {author} {\bibfnamefont {A.}~\bibnamefont
  {{Paranjape}}}, \bibinfo {author} {\bibfnamefont {O.}~\bibnamefont
  {{Hahn}}},\ and\ \bibinfo {author} {\bibfnamefont {R.~K.}\ \bibnamefont
  {{Sheth}}},\ }\bibfield  {title} {\bibinfo {title} {{Excursion set peaks: the
  role of shear}},\ }\href@noop {} {\bibfield  {journal} {\bibinfo  {journal}
  {arXiv e-prints}\ ,\ \bibinfo {eid} {arXiv:1611.03619}} (\bibinfo {year}
  {2016})},\ \Eprint {https://arxiv.org/abs/1611.03619} {arXiv:1611.03619
  [astro-ph.CO]} \BibitemShut {NoStop}%
\bibitem [{\citenamefont {Paranjape}\ \emph {et~al.}(2013)\citenamefont
  {Paranjape}, \citenamefont {Sheth},\ and\ \citenamefont
  {Desjacques}}]{Paranjape2013}%
  \BibitemOpen
  \bibfield  {author} {\bibinfo {author} {\bibfnamefont {A.}~\bibnamefont
  {Paranjape}}, \bibinfo {author} {\bibfnamefont {R.~K.}\ \bibnamefont
  {Sheth}},\ and\ \bibinfo {author} {\bibfnamefont {V.}~\bibnamefont
  {Desjacques}},\ }\bibfield  {title} {\bibinfo {title} {Excursion set peaks: a
  self-consistent model of dark halo abundances and clustering},\ }\href
  {https://doi.org/10.1093/mnras/stt267} {\bibfield  {journal} {\bibinfo
  {journal} {Monthly Notices of the Royal Astronomical Society}\ }\textbf
  {\bibinfo {volume} {431}},\ \bibinfo {pages} {1503–1512} (\bibinfo {year}
  {2013})}\BibitemShut {NoStop}%
\bibitem [{\citenamefont {{Bond}}\ and\ \citenamefont
  {{Myers}}(1996{\natexlab{b}})}]{Bond&Myers19962}%
  \BibitemOpen
  \bibfield  {author} {\bibinfo {author} {\bibfnamefont {J.~R.}\ \bibnamefont
  {{Bond}}}\ and\ \bibinfo {author} {\bibfnamefont {S.~T.}\ \bibnamefont
  {{Myers}}},\ }\bibfield  {title} {\bibinfo {title} {{The Peak-Patch Picture
  of Cosmic Catalogs. II. Validation}},\ }\href
  {https://doi.org/10.1086/192268} {\bibfield  {journal} {\bibinfo  {journal}
  {The Astrophysical Journal Supplement Series}\ }\textbf {\bibinfo {volume}
  {103}},\ \bibinfo {pages} {41} (\bibinfo {year}
  {1996}{\natexlab{b}})}\BibitemShut {NoStop}%
\bibitem [{\citenamefont {Stein}\ \emph {et~al.}(2018)\citenamefont {Stein},
  \citenamefont {Alvarez},\ and\ \citenamefont {Bond}}]{Stein2018}%
  \BibitemOpen
  \bibfield  {author} {\bibinfo {author} {\bibfnamefont {G.}~\bibnamefont
  {Stein}}, \bibinfo {author} {\bibfnamefont {M.~A.}\ \bibnamefont {Alvarez}},\
  and\ \bibinfo {author} {\bibfnamefont {J.~R.}\ \bibnamefont {Bond}},\
  }\bibfield  {title} {\bibinfo {title} {The mass-peak patch algorithm for fast
  generation of deep all-sky dark matter halo catalogues and itsn-body
  validation},\ }\href {https://doi.org/10.1093/mnras/sty3226} {\bibfield
  {journal} {\bibinfo  {journal} {Monthly Notices of the Royal Astronomical
  Society}\ }\textbf {\bibinfo {volume} {483}},\ \bibinfo {pages} {2236–2250}
  (\bibinfo {year} {2018})}\BibitemShut {NoStop}%
\bibitem [{\citenamefont {Glorot}\ and\ \citenamefont
  {Bengio}(2010)}]{glorot10}%
  \BibitemOpen
  \bibfield  {author} {\bibinfo {author} {\bibfnamefont {X.}~\bibnamefont
  {Glorot}}\ and\ \bibinfo {author} {\bibfnamefont {Y.}~\bibnamefont
  {Bengio}},\ }\bibfield  {title} {\bibinfo {title} {Understanding the
  difficulty of training deep feedforward neural networks}\ }(\bibinfo
  {publisher} {JMLR Workshop and Conference Proceedings},\ \bibinfo {address}
  {Chia Laguna Resort, Sardinia, Italy},\ \bibinfo {year} {2010})\ pp.\
  \bibinfo {pages} {249--256}\BibitemShut {NoStop}%
\bibitem [{\citenamefont {Nwankpa}\ \emph {et~al.}(2018)\citenamefont
  {Nwankpa}, \citenamefont {Ijomah}, \citenamefont {Gachagan},\ and\
  \citenamefont {Marshall}}]{Nwankpa2018}%
  \BibitemOpen
  \bibfield  {author} {\bibinfo {author} {\bibfnamefont {C.}~\bibnamefont
  {Nwankpa}}, \bibinfo {author} {\bibfnamefont {W.}~\bibnamefont {Ijomah}},
  \bibinfo {author} {\bibfnamefont {A.}~\bibnamefont {Gachagan}},\ and\
  \bibinfo {author} {\bibfnamefont {S.}~\bibnamefont {Marshall}},\ }\bibfield
  {title} {\bibinfo {title} {Activation functions: Comparison of trends in
  practice and research for deep learning.},\ }\href@noop {} {\bibfield
  {journal} {\bibinfo  {journal} {arXiv}\ ,\ \bibinfo {pages} {Preprint at
  arXiv:1811.03378}} (\bibinfo {year} {2018})}\BibitemShut {NoStop}%
\bibitem [{\citenamefont {Denil}\ \emph {et~al.}(2013)\citenamefont {Denil},
  \citenamefont {Shakibi}, \citenamefont {Dinh}, \citenamefont {Ranzato},\ and\
  \citenamefont {de~Freitas}}]{Denil2013}%
  \BibitemOpen
  \bibfield  {author} {\bibinfo {author} {\bibfnamefont {M.}~\bibnamefont
  {Denil}}, \bibinfo {author} {\bibfnamefont {B.}~\bibnamefont {Shakibi}},
  \bibinfo {author} {\bibfnamefont {L.}~\bibnamefont {Dinh}}, \bibinfo {author}
  {\bibfnamefont {M.}~\bibnamefont {Ranzato}},\ and\ \bibinfo {author}
  {\bibfnamefont {N.}~\bibnamefont {de~Freitas}},\ }\bibfield  {title}
  {\bibinfo {title} {Predicting parameters in deep learning},\ }in\ \href@noop
  {} {\emph {\bibinfo {booktitle} {Proceedings of the 26th International
  Conference on Neural Information Processing Systems - Volume 2}}},\ \bibinfo
  {series and number} {NIPS'13}\ (\bibinfo  {publisher} {Curran Associates
  Inc.},\ \bibinfo {address} {Red Hook, NY, USA},\ \bibinfo {year} {2013})\ p.\
  \bibinfo {pages} {2148–2156}\BibitemShut {NoStop}%
\bibitem [{\citenamefont {Scardapane}\ \emph {et~al.}(2017)\citenamefont
  {Scardapane}, \citenamefont {Comminiello}, \citenamefont {Hussain},\ and\
  \citenamefont {Uncini}}]{Scardapane2017}%
  \BibitemOpen
  \bibfield  {author} {\bibinfo {author} {\bibfnamefont {S.}~\bibnamefont
  {Scardapane}}, \bibinfo {author} {\bibfnamefont {D.}~\bibnamefont
  {Comminiello}}, \bibinfo {author} {\bibfnamefont {A.}~\bibnamefont
  {Hussain}},\ and\ \bibinfo {author} {\bibfnamefont {A.}~\bibnamefont
  {Uncini}},\ }\bibfield  {title} {\bibinfo {title} {Group sparse
  regularization for deep neural networks},\ }\href@noop {} {\bibfield
  {journal} {\bibinfo  {journal} {Neurocomputing}\ }\textbf {\bibinfo {volume}
  {241}},\ \bibinfo {pages} {81} (\bibinfo {year} {2017})}\BibitemShut
  {NoStop}%
\end{thebibliography}%

\end{document}